\documentclass[
    twocolumn,
    prx
]{revtex4-2}
\usepackage{zx_config}
\begin{document}

\title{Application of ZX-calculus to Quantum Architecture Search} 

\author{Tom Ewen\,\orcidlink{0009-0007-4028-6698}}
\email[]{tom.ewen@itwm.fraunhofer.de}
\affiliation{Fraunhofer ITWM}
\author{Ivica Turkalj}
\email[]{ivica.turkalj@itwm.fraunhofer.de}
\affiliation{Fraunhofer ITWM}
\author{Patrick Holzer\,\orcidlink{0009-0004-2975-8365}}
\email[]{patrick.holzer@itwm.fraunhofer.de}
\affiliation{Fraunhofer ITWM}
\author{Mark-Oliver Wolf\,\orcidlink{0000-0002-3698-9266}}
\email[]{mark-oliver.wolf@itwm.fraunhofer.de}
\affiliation{Fraunhofer ITWM}

\date{\today}
\clearpage
\begin{abstract}
This paper presents a novel approach to quantum architecture search by integrating the techniques of ZX-calculus with Genetic Programming (GP) to optimize the structure of parameterized quantum circuits employed in Quantum Machine Learning (QML).
Recognizing the challenges in designing efficient quantum circuits for QML, we propose a GP framework that utilizes mutations defined via ZX-calculus, a graphical language that can simplify visualizing and working with quantum circuits.
Our methodology focuses on evolving quantum circuits with the aim of enhancing their capability to approximate functions relevant in various machine learning tasks.
We introduce several mutation operators inspired by the transformation rules of ZX-calculus and investigate their impact on the learning efficiency and accuracy of quantum circuits.
The empirical analysis involves a comparative study where these mutations are applied to a diverse set of quantum regression problems, measuring performance metrics such as the percentage of valid circuits after the mutation, improvement of the objective, as well as circuit depth and width.
Our results indicate that certain ZX-calculus-based mutations perform significantly better than others for Quantum Architecture Search (QAS) in all metrics considered.
They suggest that ZX-diagram based QAS results in shallower circuits and more uniformly allocated gates than crude genetic optimization based on the circuit model.
\end{abstract}
\maketitle 

\section{Introduction}\label{sec:introduction}
Quantum Machine Learning (QML) is a highly active field of research with applications in various fields~\cite{biamonteQuantumMachineLearning2017}.
The performance of QML methods heavily depends on the underlying Parameterized Quantum Circuits (PQC) used~\cite{benedettiParameterizedQuantumCircuits2019}.
PQCs with an adjustable structure are a promising candidate for exploiting quantum computers on Noisy Intermediate Scale-Quantum (NISQ) devices~\cite{ostaszewskiStructureOptimizationParameterized2021,grimsleyAdaptiveVariationalAlgorithm2019,huangRobustResourceefficientQuantum2022,bilkisSemiagnosticAnsatzVariable2023}.

If a particular problem is to be solved by a PQC, its structure must fulfill several, contradictory requirements.
On the one hand, it has to be sufficiently expressive to model the problem to be solved~\cite{holzerSpectralInvarianceMaximality2024}.
On the other hand, it needs to be executable on a NISQ device, meaning it needs to be shallow and not use too many two qubit gates~\cite{haugCapacityQuantumGeometry2021,duLearnabilityQuantumNeural2021}.
The process of finding well suited PQCs has, in recent years, been discussed in the literature under the umbrella term Quantum Architecture Search (QAS)~\cite{duQuantumCircuitArchitecture2022, kuoQuantumArchitectureSearch2021a, zhangDifferentiableQuantumArchitecture2022}.

One approach to QAS is the paradigm of Genetic Programming (GP), which has been used in Quantum Computing~\cite{toulouseAutomaticQuantumComputer2006,rubinsteinEvolvingQuantumCircuits2001,langdonFoundationsGeneticProgramming2002}, and specifically in QML~\cite{kondratyevNonDifferentiableLearningQuantum2020,tangQubitADAPTVQEAdaptiveAlgorithm2021,dingMultiObjectiveEvolutionaryArchitecture2023,wolfQuantumArchitectureSearch2023}.
GP is a computational technique that iteratively evolves computer programs to solve specific problems~\cite{kozaGeneticProgrammingMeans1994}.
Inspired by the principles of biological evolution, GP simulates the process of natural selection, where the fittest individuals, i.e., programs, are chosen to reproduce.
The core mechanism of GP involves the creation of an initial population of random programs, which are then evaluated based on one or multiple fitness functions measuring their performance on the given task.
The best-performing programs are selected to undergo genetic operations such as crossover (recombining parts of two programs), mutation (randomly altering parts of a program) and replication (no changes) to result in new programs that inherit features from their parents.
Over successive generations, the population evolves, and programs that better solve the problem become more prevalent.
Popular implementations of this paradigm are the NSGA-II algorithm~\cite{debFastElitistMultiobjective2002} or Cartesian GP~\cite{millerCartesianGeneticProgramming2000}.

Rather than directly mutating the quantum circuits with GP, we follow the approach of~\cite{barnesGeneticEvolutionQuantum2020} and mutate ZX-diagrams.
ZX-diagrams together with the ZX-calculus are an alternative, graphical way of representing and working with quantum circuits~\cite{coeckeInteractingQuantumObservables2008}, that allow a reduced redundancy in comparison to quantum circuits.
This concept has, for example, been used in work on error-correction codes~\cite{chancellorGraphicalStructuresDesign2023}, measurement-based quantum computing~\cite{duncanRewritingMeasurementBasedQuantum2010} or on topology aware optimization of circuits~\cite{gogiosoAnnealingOptimisationMixed2023}.
Recently the ZX-calculus has also been used for Quantum Program Synthesis, the task to approximate a given circuit by another, ideally simpler one~\cite{barnesGeneticEvolutionQuantum2020}.

In this work, we propose to represent PQCs by parameterized ZX-diagrams instead of PQCs and apply Genetic QAS on these diagrams.
We collect possible mutations from the literature and extend the list by some new ideas.
We perform numerical studies to evaluate and compare these mutations.

This article is structured as follows.
We start with a general introduction to the ZX-calculus in~\cref{sec:zx_calculus} and demonstrate how it can be used for genetic architecture search in~\cref{sec:application_zx_qas}.
In~\cref{sec:experiments} we evaluate numerical experiments we performed to compare different mutation strategies and benchmark the QAS with ZX-diagrams against the version with circuit based mutations.
Finally, we conclude and give an outlook in~\cref{sec:conclusion}.

\section{ZX-calculus}\label{sec:zx_calculus}
In this section we give a short introduction to ZX-calculus.
We will focus on those aspects that we will need for the application to QAS.
Our discussion closely follows the presentation in~\cite{duncanGraphtheoreticSimplificationQuantum2020,kissingerReducingNumberNonClifford2020,backensThereBackAgain2021}.
For a more detailed introduction, please refer to~\cite{vandeweteringZXcalculusWorkingQuantum2020,coeckePicturingQuantumProcesses2017,coeckeQuantumPictures2022}.

For the purpose of this paper, it is sufficient to consider ZX-diagrams as a graphical 
notation for linear maps (acting between vector spaces with tensor product structure) 
that comes with a set of useful transformation rules (called ZX-calculus).
A more rigorous discussion of diagrams in terms of formal languages
and category theory can be found in the relevant literature~\cite{hazewinkelHandbookAlgebraVolume1996,piedeleuIntroductionStringDiagrams2023,hinzeIntroducingStringDiagrams2023}.

Let \( \B \) denote the two-dimensional Hilbert space \( \C^2 \).
For \( n \in \N \) let \( \Bn \) be the \(n\)-fold tensor product of \( \B \).
The set \( \N \) will always include \(0\).

A useful property of ZX-diagrams is that a few basic diagrams are sufficient to describe any linear map
between \(\Bn \) and \(\Bm \).
Let \(n,m \in \N \) and \(\alpha \in \R \).
The linear map 
\begin{align*}
    \underbrace{\ket{0\ldots0}}_{m} \underbrace{\bra{0\ldots0}}_{n} 
    + e^{i \alpha} \underbrace{\ket{1\ldots1}}_{m} \underbrace{\bra{1\ldots1}}_{n}
\end{align*}
from \(\Bn \) to \(\Bm \) is depicted as
\begin{align*}
    \begin{ZX}
        \leftManyDots{n} \zxZ{\alpha} \rightManyDots{m}
    \end{ZX}
\end{align*}
and the diagram is called \emph{\(Z\)-spider}.

The linear map
\begin{align*}
    \underbrace{\ket{+\cdots+}}_{m}\underbrace{\bra{+\cdots+}}_{n} 
    + e^{i \alpha} \underbrace{\ket{-\cdots-}}_{m}\underbrace{\bra{-\cdots-}}_{n}
\end{align*}
from \( \Bn \) to \( \Bm \) is depicted as
\begin{align*}
    \begin{ZX}
        \leftManyDots{n} \zxX{\alpha} \rightManyDots{m}
    \end{ZX}
\end{align*}
and the diagram is called \emph{\(X\)-spider}.

Two spiders can be composed in sequence by joining the output wires of the first
with some input wires of the second. The corresponding operation for maps is the composition
of linear maps. Two spiders can also be composed in parallel by stacking them on top of each other.
This corresponds to the tensor product of linear maps. 
A collection of spiders that are composed in sequence or in parallel is called a \emph{ZX-diagram}.
Note that the cases \(n=0\) or \(m=0\) are included. In this case, \( \B^{\otimes 0}=\C \) by definition.

The following ZX-diagram, since the corresponding linear map is given by the Hadamard gate, is called a \emph{Hadamard-diagram}:
\begin{equation*}
    \begin{ZX}
        \zxN{} \rar & [\zxwCol] \zxFracZ{\pi}{2} \ar[r] & \zxFracX{\pi}{2} \ar[r] 
        & \zxFracZ{\pi}{2} \rar & [\zxwCol] \zxN{}
    \end{ZX}
\end{equation*}
Because this diagram often occurs as part of larger diagrams, 
two simplified notations are introduced:
\begin{equation*}
    \begin{ZX}
        \zxN{} \rar & [\zxwCol] \zxH{} \rar & [\zxwCol] \zxN{}
    \end{ZX}
        := 
    \begin{ZX}
        \zxN{} \rar & [\zxwCol] \zxFracZ{\pi}{2} \ar[r] & \zxFracX{\pi}{2} \ar[r] 
        & \zxFracZ{\pi}{2} \rar & [\zxwCol] \zxN{}
    \end{ZX}
\end{equation*}
\begin{equation*}
    \begin{ZX}
        \leftManyDots{} \zxZ{} \ar[r,blue] & [\zxwCol] \zxN{} \ar[r,blue] & [\zxwCol] \zxN{} \ar[r,blue] 
        & [\zxwCol] \zxZ{} \rightManyDots{}
    \end{ZX}
    :=
    \begin{ZX}
        \leftManyDots{} \zxZ{} \rar & [\zxwCol] \zxH{} \rar & [\zxwCol] \zxZ{} \rightManyDots{}
    \end{ZX}
\end{equation*}
Blue edges are also called \emph{Hadamard edges}.

A \emph{phase gadget} is a diagram of the form
\begin{equation*}   
    \begin{ZX}
        \zxZ{\alpha} \ar[r,blue] & [\zxwCol] \zxN{} \ar[r,blue] &
        \zxZ{} \rightManyDots{} 
    \end{ZX}
\end{equation*} 
Phase gadgets are used extensively in circuit optimization~\cite{backensThereBackAgain2021,kissingerReducingNumberNonClifford2020}. 
We will use them as a method to perform mutations of ZX-diagrams.

Two ZX-diagrams are called equal, iff their induced linear maps are equal up to a 
scalar factor in \(\C \setminus \{0\} \).
One can show (\cite{coeckePicturingQuantumProcesses2017,vandeweteringZXcalculusWorkingQuantum2020}) 
that equality defined in this way is consistent with the more 
intuitive view that two diagrams are equal if one can be reshaped into the other by 
rearranging the vertices, bending, unbending, crossing, and uncrossing wires, 
while keeping the connectivity and the order of the inputs and outputs intact.

\begin{figure}
    \centering
    \includegraphics[width=\linewidth]{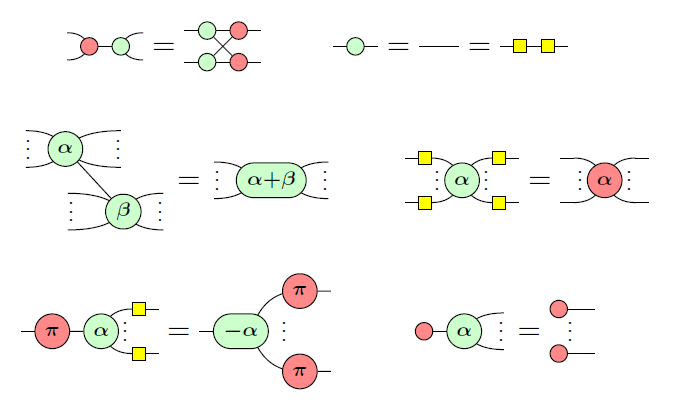}
    \caption{Rewrite rules of ZX-calculus. The equations hold for every \(\alpha, \beta \in \R \).
    They also hold if the colors are interchanged.}\label{fig:zx_calculus_clifford}
\end{figure}

ZX-diagrams fulfill a collection of equations known as the ZX-calculus. There are 
multiple variations of the ZX-calculus and the specific set of rules we will be using is shown in~\cref{fig:zx_calculus_clifford}.

A ZX-diagram is called \emph{circuit-like} if 
the number of input wires equals the number of output wires and, additionally, if
it is composed of the following diagrams:
\begin{equation*}
    \mathrm{CNOT} = 
    \begin{ZX}
        \zxN{} \rar & [\zxwCol] \zxZ{} \dar \rar & [\zxwCol] \zxN{} \\
        \zxN{} \rar & \zxX{} \rar & [\zxwCol] \zxN{} \\
    \end{ZX}
    \qquad
    Z_{\alpha} = 
    \begin{ZX} 
        \zxN{} \rar &[\zxwCol] \zxZ{\alpha} \rar &[\zxwCol] \zxN{} 
    \end{ZX}
    \qquad
    H = 
    \begin{ZX}
        \zxN{} \rar & [\zxwCol] \zxH{} \rar & [\zxwCol] \zxN{}
    \end{ZX}
\end{equation*}

Since the diagram \(\begin{ZX} \zxN{} \rar &[\zxwCol] \zxZ{\alpha} \rar &[\zxwCol] \zxN{} \end{ZX}\)
can realize both the \(S\)-gate (by \(\alpha=\tfrac{\pi}{2}\)) and the \(T\)-gate 
(by \(\alpha = \tfrac{\pi}{4}\)), this family of diagrams comprises the universal Clifford+\(T\) gateset, 
hence circuit-like diagrams correspond exactly to quantum circuits.

The above discussion shows a straightforward method to
transform circuits into ZX-diagrams: express the circuit in 
terms of the Clifford+T gateset, and interpret the circuit as a circuit-like diagram. 
The circuit extraction problem deals with the question of how to transform ZX-diagrams 
into circuit-like diagrams. 
To address this question, it is necessary to introduce two new concepts (graph-like diagrams and gFlows).

A ZX-diagram is graph-like if the following conditions hold:
\begin{enumerate}
    \item Every spider in the diagram is a \(Z\)-spider.
    \item Every wire between two \(Z\)-spiders is a Hadamard edge. 
    \item There are no parallel Hadamard edges and no self-loops.
    \item Every input-wire and every output wire is connected to a \(Z\)-spider.
    \item Every \(Z\)-spider is connected to at most one input wire or output-wire.   
\end{enumerate}

It was shown in~\cite{duncanGraphtheoreticSimplificationQuantum2020}, Lemma \(3.2\),
that every ZX-diagram is equal to a graph-like ZX-diagram. 
From our point of view, the graph-like property is a special kind of normal form that 
we will use to perform mutations and to convert ZX-diagrams into circuit-like diagrams.

We denote by \(D'\) the graph-like diagram that results from a diagram \(D\) by applying the algorithm from~\cite{duncanGraphtheoreticSimplificationQuantum2020} Lemma \(3.2\).
This is also the output of the \verb|pyzx| function \verb|to_graph_like()|.

Generalized flows are a purely graph-theoretical concept that neither make use of 
the phases contained in the diagram nor of the interpretation of a ZX-diagram as a linear map.
The exact definition of gFlow is quite technical and beyond the scope of this paper. 
We are only interested in the relation between gFlows and extractability of circuits, 
which is summarized in~\cref{thm:gflow_unitary}. 
For more details on gFlows, please refer to~\cite{backensThereBackAgain2021}.

For the formulation of the theorem we further need the following notion.

An \emph{open graph} is a triple \((G,I,O)\) where \(G=(V,E)\) is an undirected graph, and \(I\subseteq V\), \(O \subseteq V\).
Vertices in \(I\) are called inputs and vertices in \(O\) are called outputs.

Let \(D\) be a ZX-diagram.
The \emph{underlying open graph} of \(D\), denoted \(G(D)\) is defined as follows:
vertices are the spiders of \(D\), edges in the graph are Hadamard edges in \(D\), \(I\) consists of spiders connected (not necessarily by a Hadamard edge) to an input wire, and \(O\) consist of spiders connected (not necessarily by a Hadamard edge) to an output wire.
\begin{theorem}\label{thm:gflow_unitary}\ 
    \begin{enumerate}[label=\alph*)]
        \item Let \(D\) be a circuit-like ZX-diagram and \(D'\) the corresponding graph-like diagram. Then \(G(D')\) admits a gFlow. 
        \item Let \(D'\) be a graph-like diagram. If \(D'\) has the same number of input and output wires, and if the underlying open graph \(G(D')\) has a gFlow, then the linear map associated to \(D'\) is unitary.
    \end{enumerate}
\end{theorem}
For the proof we refer to~\cite{duncanGraphtheoreticSimplificationQuantum2020,backensThereBackAgain2021}.
Section \(7\) in~\cite{duncanGraphtheoreticSimplificationQuantum2020} presents an algorithm which
transforms a graph-like diagram \(D'\) into a circuit-like diagram, given that \(G(D')\) has a gFlow. 
An extended version can be found in~\cite{backensThereBackAgain2021}.
The latter is available in \verb|pyzx| 
in form of the function \verb|extract_circuit()|.

If one is interested in the extractability of circuits, then~\cref{thm:gflow_unitary} 
motivates the consideration of diagram transformations that preserve the gFlow property.
We will make use of such transformations when specifying our mutations in the next section.
\section{Application of ZX-calculus to Quantum Architecture Search}\label{sec:application_zx_qas}

In the current NISQ era it is a common approach for QML to use PQCs and optimize the parameters with a classical optimizer~\cite{biamonteQuantumMachineLearning2017}.
We want to adopt this procedure for parameterized ZX-diagrams.

For the implementation of parameterized ZX-diagrams we use a combination of the python libraries \verb|pyzx|~\cite{kissingerPyZXLargeScale2020}, \verb|sympy|~\cite{10.7717/peerj-cs.103}, \verb|pennylane|~\cite{bergholmPennyLaneAutomaticDifferentiation2022a} and \verb|jax|~\cite{deepmind2020jax}.
We use \verb|sympy| symbols as the phase values in the ZX-diagrams, these diagrams then can be converted to \verb|pennylane| circuits on which we can very efficiently optimize parameters by automated differentiation utilizing \verb|jax|.
This allows us to apply diagram manipulations on parameterized diagrams while still being able to efficiently calculate gradients regarding the parameters.

In principle, we could avoid the usage of quantum simulators by either using gradient free optimizers, like SPSA~\cite{spallOverviewSimultaneousPerturbation1998}, or use parameter shift rules for the gradients~\cite{schuldEvaluatingAnalyticGradients2019}.
But both alternatives slow down the calculations and deliver worse results, which both is decremental to the goal of this work, to analyze the potential of the ZX-calculus for QAS.

\subsection{Mutations}
We will now take a closer look at the mutations that were used in our algorithm.
For an open graph \((G,I,O)\)
we use the abbreviated notation \(\nO := V\setminus O\) for non-output vertices
and \(\nI = V \setminus I\) for non-input vertices. The neighborhood of 
\(v \in V\) in \(G\) is denoted
by \(N_G(v)\). We write \(v \sim w\) if two vertices \(v,w \in  V\) are connected by an edge.

\subsubsection*{Local Complementation~--~\texorpdfstring{\(M_1\)}{M1}}
The first mutation we describe is based on the following concept from graph theory.

Let \(G = (V,E)\) be a graph and \(u \in V\). The \emph{local complementation} of \(G\) at \(u\),
denoted \( G\star u\), is the graph formed by complementing the subgraph \(N_G(u)\).
Thus, \(G\) is the same graph as \(G \star u\), except on \(N_G(u)\), where we have:
\(\forall \thickspace v,w \in N_G(u): (v,w) \in E' \iff (v,w) \notin E\).

Now, let \(D'\) be a graph-like diagram and \(G=G(D')\) its underlying open graph.
Since \(D'\) is graph-like, there is a one to one correspondence, both between vertices in \(G\) and spiders in \(D'\) and between edges in \(G\) and Hadamard edges in \(D'\).
To perform the mutation, we pick a vertex \(u \notin I \cup O\) from \(G\) at random and consider the local complementation \((G \star u)\setminus \{u\} \).
We then apply the analog transformation on the corresponding spiders and Hadamard edges in \(D'\).
The motivation for this mutation is based on the next theorem.
See the appendix of~\cite{duncanGraphtheoreticSimplificationQuantum2020} for the proof.
\begin{theorem}\label{thm:local_complementation}
    Let \((G,I,O)\) be an open graph with gFlow. Then, for all \(u \notin I \cup O\),
    \(((G \star u)\setminus \{u\},I,O)\) is an open graph with gFlow.
\end{theorem}
It is important to mention that we do not change the phases of the spiders involved. 
In general, this has the consequence that the mutated ZX-diagram represents a new linear map, 
which is unitary if \(D'\) was representing a unitary 
(due to~\cref{thm:local_complementation,thm:gflow_unitary}).
\begin{equation*}
    \begin{ZX}
        \zxN{} \rar 
        & [\zxwCol] \zxZ{\alpha_1} \ar[r,blue] 
        & [\zxwCol] \zxN{} \ar[r,blue] & [\zxwCol] \zxZ{\alpha_u} \ar[r,blue] \ar[d,blue]
        & [\zxwCol] \zxN{} \ar[r,blue] & [\zxwCol] \zxZ{\alpha_2} \rar 
        & [\zxwCol] \zxN{}
        \\
        \zxN{} \rar 
        & [\zxwCol] \zxZ{\alpha_3} \ar[r,blue] 
        & [\zxwCol] \zxN{} \ar[r,blue] & [\zxwCol] \zxZ{\alpha_4} \ar[r,blue]
        & [\zxwCol] \zxN{} \ar[r,blue] & [\zxwCol] \zxZ{\alpha_5} \rar 
        & [\zxwCol] \zxN{}
    \end{ZX}
    \curly
    \begin{ZX}
        \zxN{} \rar 
        & [\zxwCol] \zxZ{\alpha_1} \ar[r,blue] \ar[rrd,blue]
        & [\zxwCol] \zxN{} \ar[r,blue] & [\zxwCol] \zxN{} \ar[r,blue] 
        & [\zxwCol] \zxN{} \ar[r,blue] & [\zxwCol] \zxZ{\alpha_2} \rar 
        & [\zxwCol] \zxN{}
        \\
        \zxN{} \rar 
        & [\zxwCol] \zxZ{\alpha_3} \ar[r,blue] 
        & [\zxwCol] \zxN{} \ar[r,blue] & [\zxwCol] \zxZ{\alpha_4} \ar[r,blue] \ar[rru,blue]
        & [\zxwCol] \zxN{} \ar[r,blue] & [\zxwCol] \zxZ{\alpha_5} \rar 
        & [\zxwCol] \zxN{}
    \end{ZX}  
\end{equation*}
The above is an example, if \(u\) corresponds to the spider with phase \(\alpha_u\).

\subsubsection*{Inverse Local Complementation~--~\texorpdfstring{\(M_2\)}{M2}}
Inverse local complementation refers to the transformation that corresponds to the
reverse direction in the figure above. This transformation introduces a new spider into the diagram.

From the set of inner spiders, candidates are sampled which will belong to the neighborhood 
of the new spider. 
First, the number of spiders in the potential neighborhood is sampled
according to some distribution. Then the corresponding number of spiders is drawn 
from the set of inner spiders.
The phase of the new spider is either a trainable parameter or a parameter for data input,
which is determined randomly.

This transformation is generally not gFlow preserving.
\subsubsection*{Pivoting~--~\texorpdfstring{\(M_3\)}{M3}}
The pivoting mutation is based on a repeated application of local complementation. 
In more detail, we consider the following transformation of graphs.

Let \(G=(V,E)\) be a graph and \(u,v \in V\) with \(u \sim v\). The \emph{pivot}
of \(G\) at \(uv\), denoted \(G \wedge uv\), is defined as \(((G \star u) \star v) \star u\).
The transformation can be understood as follows.
Let \( U := N_G(u) \setminus N_G(v), V := N_G(v) \setminus N_G(u)\) and \(W := N_G(u) \cap N_G(v)\).
The new graph \((G \wedge uv)\setminus \{u,v\} \) is constructed by removing the nodes \(u\) 
and \(v\), whereby the incidence 
relation between the subgraphs \(U,V\) and \(W\) is chosen to be complementary.

Now, let \(D'\) and \(G\) be as above. We randomly pick two connected vertices \(u,v \in I \cup O\)
and consider \((G \wedge uv)\setminus \{u,v\} \). The mutation consists of performing the corresponding
transformation on the graph-like diagram \(D'\), without making any changes to phases.
The next figure illustrates this mutation.

\begin{equation*}
    \begin{ZX}
        \zxN{} 
        & [\zxwCol] \zxN{} 
        & [\zxwCol] \zxZ{\alpha_u} \ar[r,blue] 
        & [\zxwCol] \zxN{} \ar[r,blue] 
        & [\zxwCol] \zxZ{\alpha_v} \ar[dr,blue] \ar[ddr,blue]
        & [\zxwCol] \zxN{} & [\zxwCol] \zxN{} 
        & [\zxwCol] \zxN{}
        \\
        \zxN{} \rar 
        & [\zxwCol] \zxZ{\alpha_1} \ar[ur,blue] 
        & [\zxwCol] \zxN{} & [\zxwCol] \zxN{}
        & [\zxwCol] \zxN{}  
        & [\zxwCol] \zxZ{\alpha_2} \rar 
        & [\zxwCol] \zxN{}
        \\
        \zxN{} \rar 
        & [\zxwCol] \zxZ{\alpha_3} \ar[uur,blue] 
        & [\zxwCol] \zxN{} & [\zxwCol] \zxN{}
        & [\zxwCol] \zxN{}  
        & [\zxwCol] \zxZ{\alpha_4} \rar 
        & [\zxwCol] \zxN{}
        \\
        \zxN{}
        & [\zxwCol] \zxN{} 
        & [\zxwCol] \zxN{} 
        & [\zxwCol] \zxZ{\alpha_5} \ar[uuul,blue] \ar[uuur,blue]
        & [\zxwCol] \zxN{}  
        & [\zxwCol] \zxN{}
        & [\zxwCol] \zxN{}
    \end{ZX}
    \curly
    \begin{ZX}
        \zxN{} \rar 
        & [\zxwCol] \zxZ{\alpha_1} \ar[r,blue] \ar[ddrr,blue] \ar[rrrrd, blue]
        & [\zxwCol] \zxN{} \ar[r,blue] 
        & [\zxwCol] \zxN{} \ar[r,blue] 
        & [\zxwCol] \zxN{} \ar[r,blue] 
        & [\zxwCol] \zxZ{\alpha_2} \rar 
        & [\zxwCol] \zxN{}
        \\
        \zxN{} \rar 
        & [\zxwCol] \zxZ{\alpha_3} \ar[r,blue] \ar[rrd,blue] \ar[rrrru, blue]
        & [\zxwCol] \zxN{} \ar[r,blue] 
        & [\zxwCol] \zxN{} \ar[r,blue] 
        & [\zxwCol] \zxN{} \ar[r,blue]   
        & [\zxwCol] \zxZ{\alpha_4} \rar 
        & [\zxwCol] \zxN{}
        \\
        \zxN{}
        & [\zxwCol] \zxN{} 
        & [\zxwCol] \zxN{} 
        & [\zxwCol] \zxZ{\alpha_5} \ar[urr, blue] \ar[uurr, blue]
        & [\zxwCol] \zxN{}  
        & [\zxwCol] \zxN{}
        & [\zxwCol] \zxN{}
    \end{ZX} 
\end{equation*}
Our motivation for analyzing this mutation is similar to that of local complementation.
It is gFlow, but not semantic preserving.

\begin{theorem}\label{thm:pivoting}
    Let \((G,I,O)\) be an open graph with gFlow. Then, for all \(u,v \notin I \cup O\),
    with \(u \sim v\), \(((G \wedge uv)\setminus \{u,v\},I,O)\) is an open graph with gFlow.
\end{theorem}
For the proof see the appendix of~\cite{duncanGraphtheoreticSimplificationQuantum2020}. 

\subsubsection*{Phase Gadget Addition~--~\texorpdfstring{\(M_4\)}{M4}}
This mutation adds a phase gadget to the ZX-diagram.
For example:
\begin{equation*}
    \begin{ZX}
        \zxN{} \rar 
        & [\zxwCol] \zxZ{\alpha_1} \ar[r,blue] 
        & [\zxwCol] \zxN{} \ar[r,blue] & [\zxwCol] \zxZ{\alpha_2} \ar[r,blue] \ar[d,blue]
        & [\zxwCol] \zxN{} \ar[r,blue] & [\zxwCol] \zxZ{\alpha_3} \rar 
        & [\zxwCol] \zxN{}
        \\
        \zxN{} \rar 
        & [\zxwCol] \zxZ{\alpha_4} \ar[r,blue] 
        & [\zxwCol] \zxN{} \ar[r,blue] & [\zxwCol] \zxZ{\alpha_5} \ar[r,blue]
        & [\zxwCol] \zxN{} \ar[r,blue] & [\zxwCol] \zxZ{\alpha_6} \rar 
        & [\zxwCol] \zxN{}
    \end{ZX}
    \curly
    \begin{ZX}
        \zxN{}
        & [\zxwCol] \zxN{}  
        & [\zxwCol] \zxZ{\alpha} \ar[d,blue] 
        & [\zxwCol] \zxN{} 
        & [\zxwCol] \zxN{} 
        & [\zxwCol] \zxN{}
        & [\zxwCol] \zxN{}
        \\
        \zxN{} 
        & [\zxwCol] \zxN{}
        & [\zxwCol] \zxZ{} \ar[dr,blue] \ar[ddr,blue] 
        & [\zxwCol] \zxN{} 
        & [\zxwCol] \zxN{} 
        & [\zxwCol] \zxN{}
        & [\zxwCol] \zxN{}
        \\
        \zxN{} \rar 
        & [\zxwCol] \zxZ{\alpha_1} \ar[r,blue] 
        & [\zxwCol] \zxN{} \ar[r,blue] 
        & [\zxwCol] \zxZ{\alpha_2} \ar[r,blue] \ar[d,blue]
        & [\zxwCol] \zxN{} \ar[r,blue] 
        & [\zxwCol] \zxZ{\alpha_3} \rar 
        & [\zxwCol] \zxN{}
        \\
        \zxN{} \rar 
        & [\zxwCol] \zxZ{\alpha_4} \ar[r,blue] 
        & [\zxwCol] \zxN{} \ar[r,blue] 
        & [\zxwCol] \zxZ{\alpha_5} \ar[r,blue]
        & [\zxwCol] \zxN{} \ar[r,blue] 
        & [\zxwCol] \zxZ{\alpha_6} \rar 
        & [\zxwCol] \zxN{}
    \end{ZX}
\end{equation*}
To construct the phase gadget,
we first sample
the number of outgoing wires with respect to 
some distribution.
Then, we sample the corresponding number of spiders from the set of all inner spiders, 
whereby we randomly determine
whether the phase occurring in the phase gadget 
is a trainable parameter or data input. 
\subsubsection*{Edge Flipping~--~\texorpdfstring{\(M_5\)}{M5}}
Edge flipping refers to the mutation that converts a Hadamard edge into a regular edge and vice versa.
For example:
\begin{equation*}
    \begin{ZX}
        \zxN{} \rar 
        & [\zxwCol] \zxZ{\alpha_1} \ar[r,blue] 
        & [\zxwCol] \zxN{} \ar[r,blue]
        & [\zxwCol] \zxZ{\alpha_2} \ar[r,blue] 
        & [\zxwCol] \zxN{} \ar[r,blue] 
        & [\zxwCol] \zxZ{\alpha_3} \rar 
        & [\zxwCol] \zxN{}
    \end{ZX}
    \curly 
    \begin{ZX}
        \zxN{} \rar 
        & [\zxwCol] \zxZ{\alpha_1} \ar[r,blue] 
        & [\zxwCol] \zxN{} \ar[r,blue]
        & [\zxwCol] \zxZ{\alpha_2} \ar[r] 
        & [\zxwCol] \zxN{} \ar[r] 
        & [\zxwCol] \zxZ{\alpha_3} \rar 
        & [\zxwCol] \zxN{}
    \end{ZX}
\end{equation*}
The edge to be flipped is selected at random from the set of all edges 
(in particular, input and output wire are permitted).

\subsubsection*{Edge Addition~--~\texorpdfstring{\(M_6\)}{M6}}
Here, we randomly select two spiders from the set of all non-connected spiders. 
An edge is then inserted between the selected spiders. 
For example:
\begin{equation*}
    \begin{ZX}
        \zxN{} \rar 
        & [\zxwCol] \zxZ{\alpha_1} \ar[r,blue] 
        & [\zxwCol] \zxN{} \ar[r,blue] & [\zxwCol] \zxZ{\alpha_2} \ar[r,blue]
        & [\zxwCol] \zxN{} \ar[r,blue] & [\zxwCol] \zxZ{\alpha_3} \rar 
        & [\zxwCol] \zxN{}
        \\
        \zxN{} \rar 
        & [\zxwCol] \zxZ{\alpha_4} \ar[r,blue] 
        & [\zxwCol] \zxN{} \ar[r,blue] & [\zxwCol] \zxZ{\alpha_5} \ar[r,blue]
        & [\zxwCol] \zxN{} \ar[r,blue] & [\zxwCol] \zxZ{\alpha_6} \rar 
        & [\zxwCol] \zxN{}
    \end{ZX}
    \curly
    \begin{ZX}
        \zxN{} \rar 
        & [\zxwCol] \zxZ{\alpha_1} \ar[r,blue] \ar[rrrrd,blue]
        & [\zxwCol] \zxN{} \ar[r,blue] & [\zxwCol] \zxZ{\alpha_2} \ar[r,blue] 
        & [\zxwCol] \zxN{} \ar[r,blue] & [\zxwCol] \zxZ{\alpha_3} \rar 
        & [\zxwCol] \zxN{}
        \\
        \zxN{} \rar 
        & [\zxwCol] \zxZ{\alpha_4} \ar[r,blue] 
        & [\zxwCol] \zxN{} \ar[r,blue] & [\zxwCol] \zxZ{\alpha_5} \ar[r,blue]
        & [\zxwCol] \zxN{} \ar[r,blue] & [\zxwCol] \zxZ{\alpha_6} \rar 
        & [\zxwCol] \zxN{}
    \end{ZX}
\end{equation*}
It is randomly determined whether 
the edge will be a Hadamard or a regular edge.

\subsubsection*{Edge Removal~--~\texorpdfstring{\(M_7\)}{M7}}
A randomly selected wire between two spiders is removed. 
In particular, input and output wire are excluded from the selection.
One example is the figure above, read in reversed direction.

\subsubsection*{Edge Swap~--~\texorpdfstring{\(M_8\)}{M8}}
In the first step, one edge is randomly selected from the set of all edges that 
connect two spiders (in particular, input and output wire are excluded). 
In the second step, two non-connected spiders are selected at random. 
The edge from the first step is removed and the two spiders from the second step are connected.
\begin{equation*}
    \begin{ZX}
        \zxN{} \rar 
        & [\zxwCol] \zxZ{\alpha_1} \ar[r,blue] \ar[d,blue]
        & [\zxwCol] \zxN{} \ar[r,blue] & [\zxwCol] \zxZ{\alpha_2} \ar[r,blue]
        & [\zxwCol] \zxN{} \ar[r,blue] & [\zxwCol] \zxZ{\alpha_3} \rar 
        & [\zxwCol] \zxN{}
        \\
        \zxN{} \rar 
        & [\zxwCol] \zxZ{\alpha_4} \ar[r,blue] 
        & [\zxwCol] \zxN{} \ar[r,blue] & [\zxwCol] \zxZ{\alpha_5} \ar[r,blue]
        & [\zxwCol] \zxN{} \ar[r,blue] & [\zxwCol] \zxZ{\alpha_6} \rar 
        & [\zxwCol] \zxN{}
    \end{ZX}
    \curly
    \begin{ZX}
        \zxN{} \rar 
        & [\zxwCol] \zxZ{\alpha_1} \ar[d,blue] 
        & [\zxwCol] \zxN{} 
        & [\zxwCol] \zxZ{\alpha_2} \ar[r,blue] \ar[d,blue]
        & [\zxwCol] \zxN{} \ar[r,blue] & [\zxwCol] \zxZ{\alpha_3} \rar 
        & [\zxwCol] \zxN{}
        \\
        \zxN{} \rar 
        & [\zxwCol] \zxZ{\alpha_4} \ar[r,blue] 
        & [\zxwCol] \zxN{} \ar[r,blue] & [\zxwCol] \zxZ{\alpha_5} \ar[r,blue]
        & [\zxwCol] \zxN{} \ar[r,blue] & [\zxwCol] \zxZ{\alpha_6} \rar 
        & [\zxwCol] \zxN{}
    \end{ZX}
\end{equation*}
The type of edge is selected at random. 
See the figure above for an example.

\subsubsection*{Edge Split~--~\texorpdfstring{\(M_9\)}{M9}}
Here, a new \(Z\)-spider is introduced on an existing edge, splitting the edge into parts.
\begin{equation*}
    \begin{ZX}
        \zxN{} \rar 
        & [\zxwCol] \zxZ{\alpha_1} \ar[r,blue] 
        & [\zxwCol] \zxN{} \ar[r,blue] & [\zxwCol] \zxZ{\alpha_2} \ar[r]
        & [\zxwCol] \zxN{}
        \\
        \zxN{} \rar 
        & [\zxwCol] \zxZ{\alpha_3} \ar[r,blue] 
        & [\zxwCol] \zxN{} \ar[r,blue] & [\zxwCol] \zxZ{\alpha_3} \ar[r] 
        & [\zxwCol] \zxN{}
    \end{ZX}
    \curly
    \begin{ZX}
        \zxN{} \rar 
        & [\zxwCol] \zxZ{\alpha_1} \ar[r,blue]
        & [\zxwCol] \zxN{} \ar[r,blue] 
        & [\zxwCol] \zxZ{\alpha} \ar[r,blue]  
        & [\zxwCol] \zxN{} \ar[r,blue] 
        & [\zxwCol] \zxZ{\alpha_2} \rar 
        & [\zxwCol] \zxN{}
        \\
        \zxN{} \rar 
        & [\zxwCol] \zxZ{\alpha_3} \ar[r,blue] 
        & [\zxwCol] \zxN{} \ar[r,blue]
        & [\zxwCol] \zxN{} \ar[r,blue] 
        & [\zxwCol] \zxN{} \ar[r,blue] 
        & [\zxwCol] \zxZ{\alpha_4} \ar[r]
        & [\zxwCol] \zxN{}
    \end{ZX}
\end{equation*}
The edge to be split is selected at random from the set of all edges. 
On the edge, the phase of the spider that is inserted is either a trainable parameter or 
a phase of input-type, randomly determined.
The additional edge created by the introduction of the new spider is of the same type 
as the edge selected at the beginning.

\subsection{Genetic Algorithm}
Our implementation of a genetic algorithm to find good performing ZX-diagrams is depicted in~\cref{alg:genetic_algorithm}.
Good performance is measured via multiple fitness criteria, for example the approximation error or the circuit depth.
\begin{figure}
\begin{algorithm}[H]
    \caption{Genetic ZX-diagram search}\label{alg:genetic_algorithm}
    \begin{algorithmic}[1]
        \State{} Generate initial ZX-diagrams.\label{alg:step_1}
        \Repeat{}
            \State{} Select ZX-diagrams to mutate.\label{alg:step_2}
            \State{} Mutate each selected ZX-diagram.\label{alg:step_3}
            \State{} Select ZX-diagrams to keep.\label{alg:step_4}
        \Until{Termination criteria reached.}
    \end{algorithmic}
\end{algorithm}
\end{figure}

It consists of the following steps.
\begin{enumerate}
    \item We initialize the ZX-diagrams, either randomly by using methods provided by the \verb|pyzx| library~\cite{kissingerPyZXLargeScale2020} (\verb|CNOT_HAD_PHASE_circuit|, \verb|cliffords|, \verb|cnot|) and then manually replacing fixed phases by parameterized ones, or by using a fixed circuit structure, that we transform into a ZX-diagram.
    \item For the selection of the diagrams, that we will mutate, we use a combination of elitism and random sampling.
    As long as the set of diagrams is smaller than the specified number of diagrams that should be mutated in every step, we take every diagram and randomly select additional diagrams from this set to reach the desired number.
    If the set of diagrams is larger, then we first take every diagram that is optimal in one of the fitness values we consider, and sample the rest randomly.
    \item Every possible type of mutation has an assigned probability with which it is performed.
    The parameters of each mutation, like the edges and vertices it is performed on, are chosen randomly.
    As the success of a mutation is not guaranteed, we try this multiple times until a mutation is successful or a given maximal number of trials is reached.
    A successful mutation is one, that produces a diagram that can be transformed into a circuit.
    It is possible that multiple mutations are performed in one step.
    \item In this step, we select the models to be used for the next loop.
    We take all the original diagrams, and the mutated diagrams together and calculate the fitness values for each.
    Then we exclude every diagram, for which another diagram is better in at least one fitness value without being worse in one of the others, i.e., we select only the non dominated diagrams.
\end{enumerate}

\section{Experiments}\label{sec:experiments}
In this section we will have a look at numerical experiments to evaluate different mutations and exemplary compare the results of QAS based on ZX-diagrams with our previous work~\cite{wolfQuantumArchitectureSearch2023} based on quantum circuits.
\subsection{Impact of different Mutations}
We are interested in two properties: how often does a mutation create a valid circuit, and how much does the mutation actually change.

The first property we verify with help of the python library \verb|pyzx|~\cite{kissingerPyZXLargeScale2020} by trying to extract a circuit.

\begin{figure}
    \centering
    \includegraphics{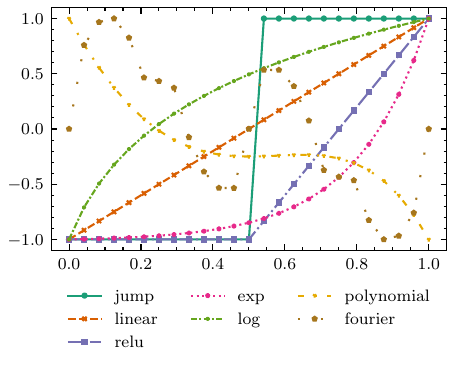}
    \caption{
        Functions used for evaluating the change introduced by a mutation.
    }\label{fig:mutation_evaluation_set}
\end{figure}

The second property is less obvious to measure.
The result of a mutation is a parameterized ZX-diagram, that represents a circuit and a unitary.
One could try to express the change in the unitary in some metric, depending on the parameters, but this will be very hard to interpret.
Instead, we look at how the performance of the ZX-diagrams changes for some example regression problems.
To make sure our results are not biased towards a specific type of problem we test different functions as benchmarks.
As seen in~\cref{fig:mutation_evaluation_set} we use a step function with a jump in the middle, the identity, a ReLU function, an exponential function, a logarithmic function, a polynomial of degree four and a weighted sum of four sinuses with different frequencies.
As our ZX-diagrams work with input in the range of zero to one and we use Pauli Z measurements, the functions are scaled accordingly.

In order to get an overview on the performance of the possible mutations we perform numerical experiments.
We generated \num{1362} random parameterized ZX-diagrams and record for each of them the number of qubits used, the number of vertices and the connectivity, which is the proportion of actual edges to the maximal possible number of edges.
Then we train the parameters for every ZX-diagram and every test function and record the training error.
The next step is applying the mutations, since all of them have a random component, we apply every mutation on every generated ZX-diagram ten times.
If the mutation was successful, meaning it resulted in a ZX-diagram that is equal to a circuit-like diagram, we train again the parameters for every test function and record the new training errors.

\begin{figure}
    \centering
    \includegraphics{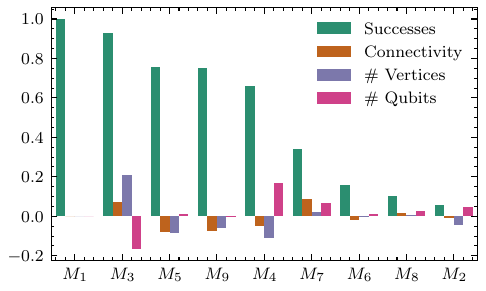}
    \caption{Parameters of the linear model for the success of a mutation. The labels on the x-axis correspond to the mutations introduced in~\cref{sec:application_zx_qas}.}\label{fig:mutation_success_rates}
\end{figure}

\begin{figure}
    \centering
    \begin{subfigure}[b]{\linewidth}
        \centering
        \includegraphics{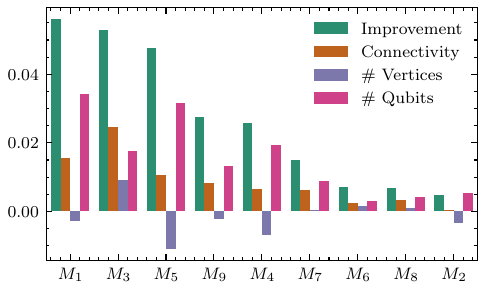}
        \caption{With unsuccessful mutations.}
    \end{subfigure}
    \begin{subfigure}[b]{\linewidth}
        \centering
        \includegraphics{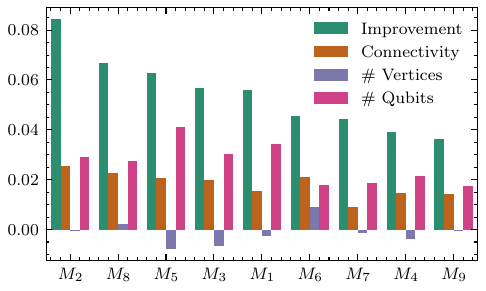}
        \caption{Only successful mutations.}
    \end{subfigure}
    \caption{Parameters of the linear model for the absolute improvement of the training error. The labels on the x-axis correspond to the mutations introduced in~\cref{sec:application_zx_qas}.}\label{fig:mutation_improvement}
\end{figure}

To analyze the performance of our genetic algorithm, we use this dataset to train three linear regression models for every type of mutation with the number of qubits, the number of vertices and the connectivity as features.
The three models differ in their target variable, for the first model we use the binary variable success or no success and for the other two we use the average absolute change in the training error.
The difference between these two is that for the first one we only consider successful mutations and for the other we assign unsuccessful mutations an improvement of zero.
We center and standardize the features to make sure the intercept equals the average of the observations and the coefficients are comparable.
The coefficients of the three models are visualized in~\cref{fig:mutation_success_rates} for the success as target and in~\cref{fig:mutation_improvement} for the improvement on the test functions.
The coefficients are plotted for every mutation individually, marked with \(M_i\), which refers to the explanations in~\cref{sec:application_zx_qas}.
The first bar is the intercept, which equals the average of the target variable over all mutations. 
For~\cref{fig:mutation_success_rates} this means it is the average success rate of the respective mutation while for~\cref{fig:mutation_improvement} the first bar represents the average absolute improvement over all generated ZX-diagrams.
The other bars represent the coefficients of the features, the second one the connectivity, then the number of vertices and finally the number of qubits.
These bars can be interpreted as how much the respective feature influences the target, we see for example that the addition of new phase gadgets (\(M_4\)) is less likely to be successful for larger number of vertices, but more likely to succeed for an increased number of qubits.

\begin{figure}
    \centering
    \includegraphics{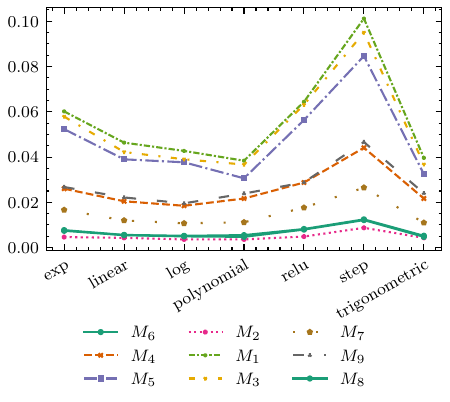}
    \caption{The average absolute improvement per mutation and test function. The labels in the legend correspond to the mutations introduced in~\cref{sec:application_zx_qas}}\label{fig:mutation_improvement_per_test_function}
\end{figure}

We further analyzed, if there is a difference in performance of the mutations for the different test functions.
In~\cref{fig:mutation_improvement_per_test_function} we can see that while the different test functions lead to different improvements, there is no evidence that some mutations work better for some test functions than for others.

After \num{122580} random mutations we come to the conclusion that while some mutations succeeded much more often than others, none have a vanishing success rate.

\subsection{Comparison of QAS with circuits and ZX-diagrams}
The next step is comparing the results of~\cref{alg:genetic_algorithm} with a similar algorithm that operates directly on quantum circuits, see~\cite{wolfQuantumArchitectureSearch2023} for more details.
Our benchmark is to find a circuit that approximates the payout of a European call option with strike \num{100} given as
\begin{equation}\label{eq:benchmark}
    f(x) = \max\left\lbrace 0, x - 100 \right\rbrace.
\end{equation}

As fitness measures we used the mean-squared-error between the prediction and the target function at \num{400} equidistant support points as measure for the quality of a model, and the circuit depth, the number of two qubit gates and the number of input encodings as a measure for the complexity of the model.

\begin{figure}
    \centering
    \includegraphics[width=\linewidth]{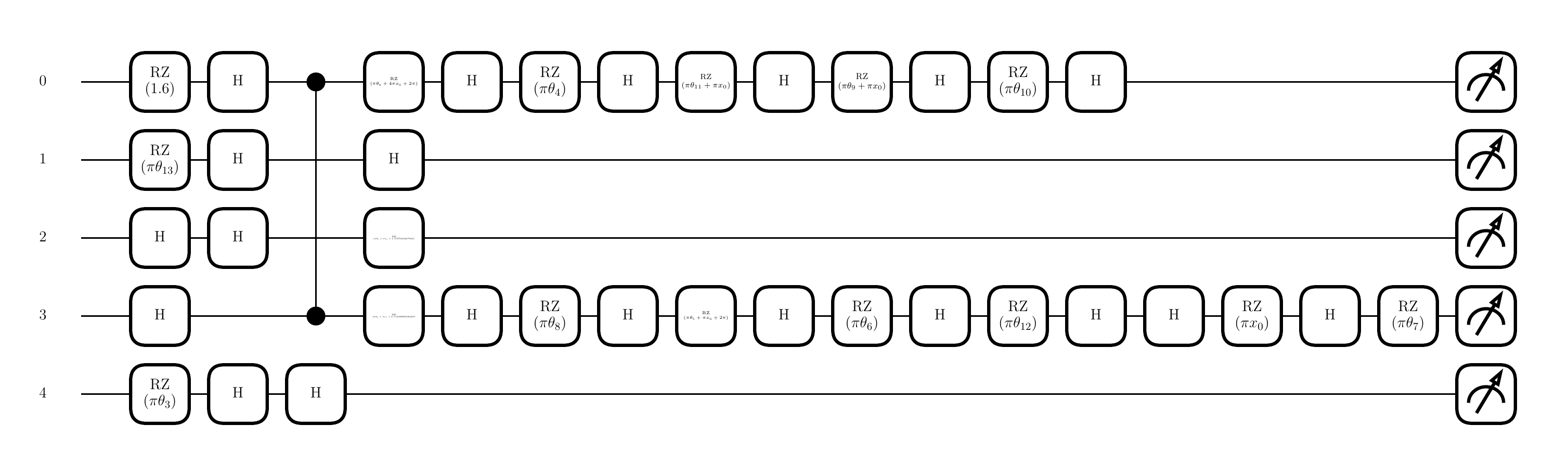}
    \caption{The circuit found by \cref{alg:genetic_algorithm} making use of ZX-diagrams for approximating \cref{eq:benchmark}. It has a depth of \num{17} uses one two qubit gate and seven input encodings and leads to a mean-square-error of \num{0.126}.}\label{fig:zx_circuit}
\end{figure}

The result of the QAS is a four dimensional Pareto front, a set of models that are non-dominated by the others.
From there we selected one model with a good trade off between the mean-squared-error and the circuit depth.
The corresponding circuit is shown in~\cref{fig:zx_circuit}.

\begin{figure}
    \centering
    \includegraphics[width=\linewidth]{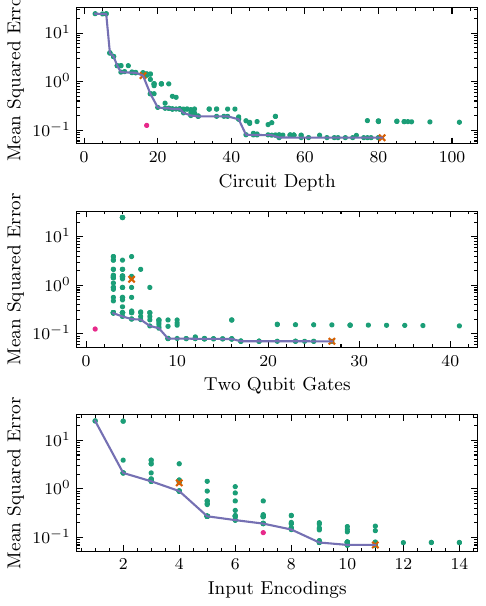}
    \caption{
        The four dimensional Pareto front of non-dominated models generated by circuit based mutations, for visualization projected onto three different planes.
        The ZX-diagram based model is included as a pink star and the models analyzed in more detail are marked with an orange cross.
    }\label{fig:circuit_pareto_front}
\end{figure}

\begin{figure}
    \centering
    \includegraphics[width=\linewidth]{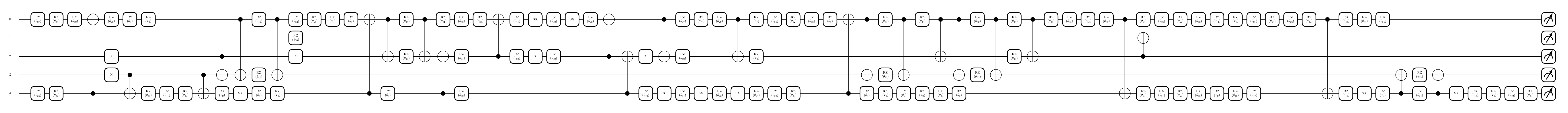}
    \caption{The circuit, found with circuit based mutations, with the lowest mean-squared-error. The detail of every gate is not readable in this scale, but also not important, the point is to show the complexity of the circuit. The circuit has a depth of \num{81} and uses \num{27} two qubit gates as well as eleven input encodings and leads to a mean-square-error of \num{0.070}.}\label{fig:circuit_lowest_error_circuit}
\end{figure}

\begin{figure}
    \centering
    \includegraphics[width=\linewidth]{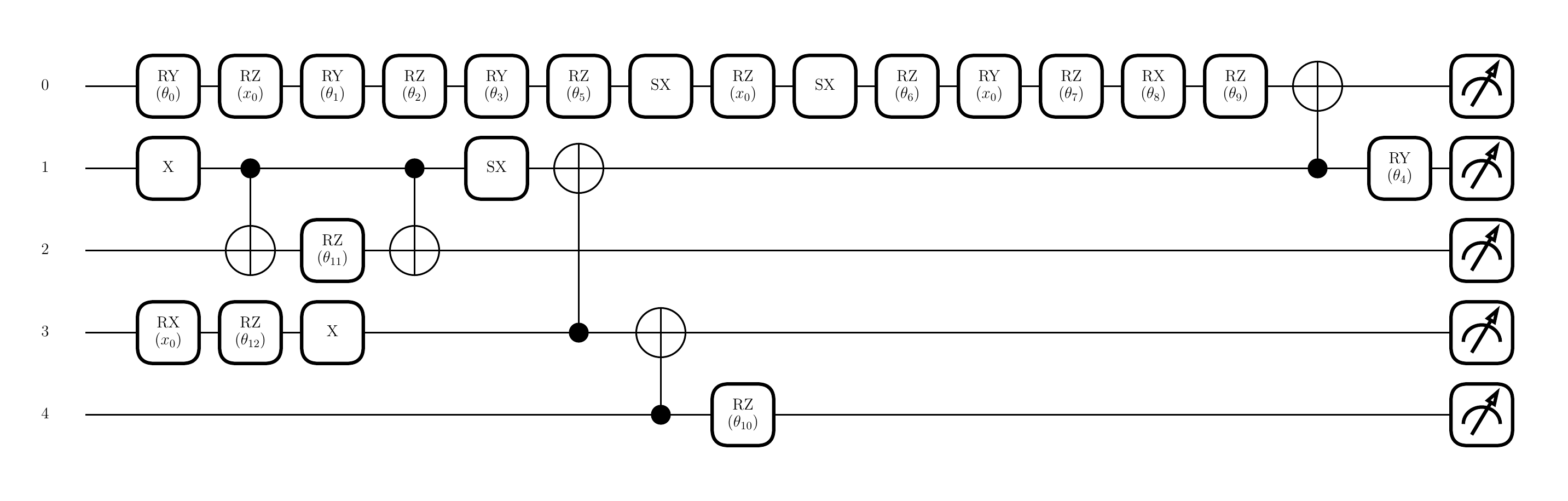}
    \caption{
        The circuit, found with circuit based mutations, with a similar circuit depth as the ZX-diagram based model. 
        The circuit has a depth of \num{16} and uses five two qubit gates as well as four input encodings and leads to a mean-square-error of \num{1.333}.
    }\label{fig:circuit_same_depth_circuit}
\end{figure}

We did the same with circuit based mutations and also obtained a Pareto front of models.
In~\cref{fig:circuit_pareto_front} we visualized the Pareto front with the selected model generated by the ZX-diagram based QAS included as Benchmark.
We can see that while there were circuit based models found that have a lower mean-squared-error than the ZX-diagram based model, they pay for that with a much higher complexity, see for example the model with the lowest mean-squared-error depicted in~\cref{fig:circuit_lowest_error_circuit}.
On the other hand, if we choose the best model which has the same or lower circuit depth as the ZX-diagram based model, we end up with a model depicted in~\cref{fig:circuit_same_depth_circuit}, which has a higher mean-squared-error than the ZX-diagram based result.
In other words, the model obtained with ZX-diagram based mutations dominates models on the Pareto front of optimal models found by circuit based mutations.

\begin{figure}
    \centering
    \includegraphics[width=\linewidth]{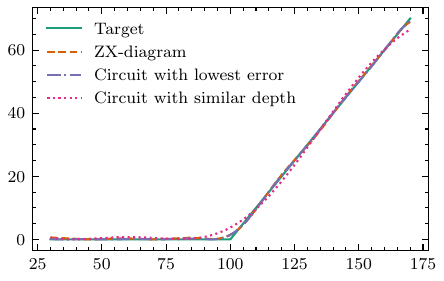}
    \caption{The approximations the different models represent. The ZX-diagram has a mean-squared-error of \num{0.126}, the circuit model with the lowest error \num{0.070} and the circuit model with the similar circuit depth \num{1.333}.}\label{fig:predictions}
\end{figure}

When comparing the predictions of the different models, see~\cref{fig:predictions}, we see that the differences are small but the ZX-diagram model and the best circuit based model, clearly perform the best.
\section{Conclusion}\label{sec:conclusion}
We extended the ideas from~\cite{barnesGeneticEvolutionQuantum2020} to PQCs and analyzed possible mutations with regard to their success probabilities and their influence on the performance on benchmark regression problems.
We analyzed mutations previously introduced in the literature and introduced new ones as well.
In numerical experiments we were able to show, that while all considered mutations have non-negligible success probabilities they have significant differences in their respective impact on our benchmark problems.
We further analyzed if and how properties of the mutated ZX-diagrams influence the performance of the mutations.

In an experimental comparison we saw indications, that the use of ZX-diagram based mutations, instead of circuit based ones, can lead to more efficient circuits for approximating functions.

With this work we want to contribute to the goal formulated in~\cite{schuldQuantumAdvantageRight2022}, moving away from imitating and trying to beat classical machine learning.
Our genetic approach in combination with the ZX-calculus can help to find new structures unique to quantum computers.

In future research, additional mutations could be introduced and compared in our benchmark setting.
More performance comparisons of \emph{classical} QAS based on the circuit structure and QAS based on ZX-diagrams are needed to solidify the indications we observed.
The interplay of this QAS technique and hardware constraints like limited amount of qubits or depth is unclear.
Furthermore, it could be analyzed if the ZX-diagram properties we observed could be included in genetic QAS algorithms to choose more fitting mutations in every step.

\begin{acknowledgments}
This work was supported by the project AnQuC-3 of the Competence Center Quantum Computing Rhineland-Palatinate (Germany).
\end{acknowledgments}
\bibliography{Bibliography}

\begin{thebibliography}{43}%
\makeatletter
\providecommand \@ifxundefined [1]{%
 \@ifx{#1\undefined}
}%
\providecommand \@ifnum [1]{%
 \ifnum #1\expandafter \@firstoftwo
 \else \expandafter \@secondoftwo
 \fi
}%
\providecommand \@ifx [1]{%
 \ifx #1\expandafter \@firstoftwo
 \else \expandafter \@secondoftwo
 \fi
}%
\providecommand \natexlab [1]{#1}%
\providecommand \enquote  [1]{``#1''}%
\providecommand \bibnamefont  [1]{#1}%
\providecommand \bibfnamefont [1]{#1}%
\providecommand \citenamefont [1]{#1}%
\providecommand \href@noop [0]{\@secondoftwo}%
\providecommand \href [0]{\begingroup \@sanitize@url \@href}%
\providecommand \@href[1]{\@@startlink{#1}\@@href}%
\providecommand \@@href[1]{\endgroup#1\@@endlink}%
\providecommand \@sanitize@url [0]{\catcode `\\12\catcode `\$12\catcode `\&12\catcode `\#12\catcode `\^12\catcode `\_12\catcode `\%12\relax}%
\providecommand \@@startlink[1]{}%
\providecommand \@@endlink[0]{}%
\providecommand \url  [0]{\begingroup\@sanitize@url \@url }%
\providecommand \@url [1]{\endgroup\@href {#1}{\urlprefix }}%
\providecommand \urlprefix  [0]{URL }%
\providecommand \Eprint [0]{\href }%
\providecommand \doibase [0]{https://doi.org/}%
\providecommand \selectlanguage [0]{\@gobble}%
\providecommand \bibinfo  [0]{\@secondoftwo}%
\providecommand \bibfield  [0]{\@secondoftwo}%
\providecommand \translation [1]{[#1]}%
\providecommand \BibitemOpen [0]{}%
\providecommand \bibitemStop [0]{}%
\providecommand \bibitemNoStop [0]{.\EOS\space}%
\providecommand \EOS [0]{\spacefactor3000\relax}%
\providecommand \BibitemShut  [1]{\csname bibitem#1\endcsname}%
\let\auto@bib@innerbib\@empty
\bibitem [{\citenamefont {Biamonte}\ \emph {et~al.}(2017)\citenamefont {Biamonte}, \citenamefont {Wittek}, \citenamefont {Pancotti}, \citenamefont {Rebentrost}, \citenamefont {Wiebe},\ and\ \citenamefont {Lloyd}}]{biamonteQuantumMachineLearning2017}%
  \BibitemOpen
  \bibfield  {author} {\bibinfo {author} {\bibfnamefont {J.}~\bibnamefont {Biamonte}}, \bibinfo {author} {\bibfnamefont {P.}~\bibnamefont {Wittek}}, \bibinfo {author} {\bibfnamefont {N.}~\bibnamefont {Pancotti}}, \bibinfo {author} {\bibfnamefont {P.}~\bibnamefont {Rebentrost}}, \bibinfo {author} {\bibfnamefont {N.}~\bibnamefont {Wiebe}},\ and\ \bibinfo {author} {\bibfnamefont {S.}~\bibnamefont {Lloyd}},\ }\bibfield  {title} {\bibinfo {title} {Quantum machine learning},\ }\href {https://doi.org/10.1038/nature23474} {\bibfield  {journal} {\bibinfo  {journal} {Nature}\ }\textbf {\bibinfo {volume} {549}},\ \bibinfo {pages} {195} (\bibinfo {year} {2017})}\BibitemShut {NoStop}%
\bibitem [{\citenamefont {Benedetti}\ \emph {et~al.}(2019)\citenamefont {Benedetti}, \citenamefont {Lloyd}, \citenamefont {Sack},\ and\ \citenamefont {Fiorentini}}]{benedettiParameterizedQuantumCircuits2019}%
  \BibitemOpen
  \bibfield  {author} {\bibinfo {author} {\bibfnamefont {M.}~\bibnamefont {Benedetti}}, \bibinfo {author} {\bibfnamefont {E.}~\bibnamefont {Lloyd}}, \bibinfo {author} {\bibfnamefont {S.}~\bibnamefont {Sack}},\ and\ \bibinfo {author} {\bibfnamefont {M.}~\bibnamefont {Fiorentini}},\ }\bibfield  {title} {\bibinfo {title} {Parameterized quantum circuits as machine learning models},\ }\href {https://doi.org/10.1088/2058-9565/ab4eb5} {\bibfield  {journal} {\bibinfo  {journal} {Quantum Science and Technology}\ }\textbf {\bibinfo {volume} {4}},\ \bibinfo {pages} {043001} (\bibinfo {year} {2019})}\BibitemShut {NoStop}%
\bibitem [{\citenamefont {Ostaszewski}\ \emph {et~al.}(2021)\citenamefont {Ostaszewski}, \citenamefont {Grant},\ and\ \citenamefont {Benedetti}}]{ostaszewskiStructureOptimizationParameterized2021}%
  \BibitemOpen
  \bibfield  {author} {\bibinfo {author} {\bibfnamefont {M.}~\bibnamefont {Ostaszewski}}, \bibinfo {author} {\bibfnamefont {E.}~\bibnamefont {Grant}},\ and\ \bibinfo {author} {\bibfnamefont {M.}~\bibnamefont {Benedetti}},\ }\bibfield  {title} {\bibinfo {title} {Structure optimization for parameterized quantum circuits},\ }\href {https://doi.org/10.22331/q-2021-01-28-391} {\bibfield  {journal} {\bibinfo  {journal} {Quantum}\ }\textbf {\bibinfo {volume} {5}},\ \bibinfo {pages} {391} (\bibinfo {year} {2021})},\ \Eprint {https://arxiv.org/abs/1905.09692} {arxiv:1905.09692 [quant-ph]} \BibitemShut {NoStop}%
\bibitem [{\citenamefont {Grimsley}\ \emph {et~al.}(2019)\citenamefont {Grimsley}, \citenamefont {Economou}, \citenamefont {Barnes},\ and\ \citenamefont {Mayhall}}]{grimsleyAdaptiveVariationalAlgorithm2019}%
  \BibitemOpen
  \bibfield  {author} {\bibinfo {author} {\bibfnamefont {H.~R.}\ \bibnamefont {Grimsley}}, \bibinfo {author} {\bibfnamefont {S.~E.}\ \bibnamefont {Economou}}, \bibinfo {author} {\bibfnamefont {E.}~\bibnamefont {Barnes}},\ and\ \bibinfo {author} {\bibfnamefont {N.~J.}\ \bibnamefont {Mayhall}},\ }\bibfield  {title} {\bibinfo {title} {An adaptive variational algorithm for exact molecular simulations on a quantum computer},\ }\href {https://doi.org/10.1038/s41467-019-10988-2} {\bibfield  {journal} {\bibinfo  {journal} {Nature Communications}\ }\textbf {\bibinfo {volume} {10}},\ \bibinfo {pages} {3007} (\bibinfo {year} {2019})}\BibitemShut {NoStop}%
\bibitem [{\citenamefont {Huang}\ \emph {et~al.}(2022)\citenamefont {Huang}, \citenamefont {Li}, \citenamefont {Hou}, \citenamefont {Wu}, \citenamefont {Yung}, \citenamefont {Bayat},\ and\ \citenamefont {Wang}}]{huangRobustResourceefficientQuantum2022}%
  \BibitemOpen
  \bibfield  {author} {\bibinfo {author} {\bibfnamefont {Y.}~\bibnamefont {Huang}}, \bibinfo {author} {\bibfnamefont {Q.}~\bibnamefont {Li}}, \bibinfo {author} {\bibfnamefont {X.}~\bibnamefont {Hou}}, \bibinfo {author} {\bibfnamefont {R.}~\bibnamefont {Wu}}, \bibinfo {author} {\bibfnamefont {M.-H.}\ \bibnamefont {Yung}}, \bibinfo {author} {\bibfnamefont {A.}~\bibnamefont {Bayat}},\ and\ \bibinfo {author} {\bibfnamefont {X.}~\bibnamefont {Wang}},\ }\bibfield  {title} {\bibinfo {title} {Robust resource-efficient quantum variational ansatz through evolutionary algorithm},\ }\href {https://doi.org/10.1103/PhysRevA.105.052414} {\bibfield  {journal} {\bibinfo  {journal} {Physical Review A}\ }\textbf {\bibinfo {volume} {105}},\ \bibinfo {pages} {052414} (\bibinfo {year} {2022})},\ \Eprint {https://arxiv.org/abs/2202.13714} {arxiv:2202.13714 [cond-mat, physics:quant-ph]} \BibitemShut {NoStop}%
\bibitem [{\citenamefont {Bilkis}\ \emph {et~al.}(2023)\citenamefont {Bilkis}, \citenamefont {Cerezo}, \citenamefont {Verdon}, \citenamefont {Coles},\ and\ \citenamefont {Cincio}}]{bilkisSemiagnosticAnsatzVariable2023}%
  \BibitemOpen
  \bibfield  {author} {\bibinfo {author} {\bibfnamefont {M.}~\bibnamefont {Bilkis}}, \bibinfo {author} {\bibfnamefont {M.}~\bibnamefont {Cerezo}}, \bibinfo {author} {\bibfnamefont {G.}~\bibnamefont {Verdon}}, \bibinfo {author} {\bibfnamefont {P.~J.}\ \bibnamefont {Coles}},\ and\ \bibinfo {author} {\bibfnamefont {L.}~\bibnamefont {Cincio}},\ }\href@noop {} {\bibinfo {title} {A semi-agnostic ansatz with variable structure for quantum machine learning}} (\bibinfo {year} {2023}),\ \Eprint {https://arxiv.org/abs/2103.06712} {arxiv:2103.06712 [quant-ph, stat]} \BibitemShut {NoStop}%
\bibitem [{\citenamefont {Holzer}\ and\ \citenamefont {Turkalj}(2024)}]{holzerSpectralInvarianceMaximality2024}%
  \BibitemOpen
  \bibfield  {author} {\bibinfo {author} {\bibfnamefont {P.}~\bibnamefont {Holzer}}\ and\ \bibinfo {author} {\bibfnamefont {I.}~\bibnamefont {Turkalj}},\ }\href {https://doi.org/10.48550/arXiv.2402.14515} {\bibinfo {title} {Spectral invariance and maximality properties of the frequency spectrum of quantum neural networks}} (\bibinfo {year} {2024}),\ \Eprint {https://arxiv.org/abs/2402.14515} {arxiv:2402.14515 [quant-ph, stat]} \BibitemShut {NoStop}%
\bibitem [{\citenamefont {Haug}\ \emph {et~al.}(2021)\citenamefont {Haug}, \citenamefont {Bharti},\ and\ \citenamefont {Kim}}]{haugCapacityQuantumGeometry2021}%
  \BibitemOpen
  \bibfield  {author} {\bibinfo {author} {\bibfnamefont {T.}~\bibnamefont {Haug}}, \bibinfo {author} {\bibfnamefont {K.}~\bibnamefont {Bharti}},\ and\ \bibinfo {author} {\bibfnamefont {M.}~\bibnamefont {Kim}},\ }\bibfield  {title} {\bibinfo {title} {Capacity and {{Quantum Geometry}} of {{Parametrized Quantum Circuits}}},\ }\href {https://doi.org/10.1103/PRXQuantum.2.040309} {\bibfield  {journal} {\bibinfo  {journal} {PRX Quantum}\ }\textbf {\bibinfo {volume} {2}},\ \bibinfo {pages} {040309} (\bibinfo {year} {2021})}\BibitemShut {NoStop}%
\bibitem [{\citenamefont {Du}\ \emph {et~al.}(2021)\citenamefont {Du}, \citenamefont {Hsieh}, \citenamefont {Liu}, \citenamefont {You},\ and\ \citenamefont {Tao}}]{duLearnabilityQuantumNeural2021}%
  \BibitemOpen
  \bibfield  {author} {\bibinfo {author} {\bibfnamefont {Y.}~\bibnamefont {Du}}, \bibinfo {author} {\bibfnamefont {M.-H.}\ \bibnamefont {Hsieh}}, \bibinfo {author} {\bibfnamefont {T.}~\bibnamefont {Liu}}, \bibinfo {author} {\bibfnamefont {S.}~\bibnamefont {You}},\ and\ \bibinfo {author} {\bibfnamefont {D.}~\bibnamefont {Tao}},\ }\bibfield  {title} {\bibinfo {title} {Learnability of {{Quantum Neural Networks}}},\ }\href {https://doi.org/10.1103/PRXQuantum.2.040337} {\bibfield  {journal} {\bibinfo  {journal} {PRX Quantum}\ }\textbf {\bibinfo {volume} {2}},\ \bibinfo {pages} {040337} (\bibinfo {year} {2021})}\BibitemShut {NoStop}%
\bibitem [{\citenamefont {Du}\ \emph {et~al.}(2022)\citenamefont {Du}, \citenamefont {Huang}, \citenamefont {You}, \citenamefont {Hsieh},\ and\ \citenamefont {Tao}}]{duQuantumCircuitArchitecture2022}%
  \BibitemOpen
  \bibfield  {author} {\bibinfo {author} {\bibfnamefont {Y.}~\bibnamefont {Du}}, \bibinfo {author} {\bibfnamefont {T.}~\bibnamefont {Huang}}, \bibinfo {author} {\bibfnamefont {S.}~\bibnamefont {You}}, \bibinfo {author} {\bibfnamefont {M.-H.}\ \bibnamefont {Hsieh}},\ and\ \bibinfo {author} {\bibfnamefont {D.}~\bibnamefont {Tao}},\ }\bibfield  {title} {\bibinfo {title} {Quantum circuit architecture search for variational quantum algorithms},\ }\href {https://doi.org/10.1038/s41534-022-00570-y} {\bibfield  {journal} {\bibinfo  {journal} {npj Quantum Information}\ }\textbf {\bibinfo {volume} {8}},\ \bibinfo {pages} {1} (\bibinfo {year} {2022})}\BibitemShut {NoStop}%
\bibitem [{\citenamefont {Kuo}\ \emph {et~al.}(2021)\citenamefont {Kuo}, \citenamefont {Fang},\ and\ \citenamefont {Chen}}]{kuoQuantumArchitectureSearch2021a}%
  \BibitemOpen
  \bibfield  {author} {\bibinfo {author} {\bibfnamefont {E.-J.}\ \bibnamefont {Kuo}}, \bibinfo {author} {\bibfnamefont {Y.-L.~L.}\ \bibnamefont {Fang}},\ and\ \bibinfo {author} {\bibfnamefont {S.~Y.-C.}\ \bibnamefont {Chen}},\ }\href {https://doi.org/10.48550/arXiv.2104.07715} {\bibinfo {title} {Quantum {{Architecture Search}} via {{Deep Reinforcement Learning}}}} (\bibinfo {year} {2021}),\ \Eprint {https://arxiv.org/abs/2104.07715} {arxiv:2104.07715 [quant-ph]} \BibitemShut {NoStop}%
\bibitem [{\citenamefont {Zhang}\ \emph {et~al.}(2022)\citenamefont {Zhang}, \citenamefont {Hsieh}, \citenamefont {Zhang},\ and\ \citenamefont {Yao}}]{zhangDifferentiableQuantumArchitecture2022}%
  \BibitemOpen
  \bibfield  {author} {\bibinfo {author} {\bibfnamefont {S.-X.}\ \bibnamefont {Zhang}}, \bibinfo {author} {\bibfnamefont {C.-Y.}\ \bibnamefont {Hsieh}}, \bibinfo {author} {\bibfnamefont {S.}~\bibnamefont {Zhang}},\ and\ \bibinfo {author} {\bibfnamefont {H.}~\bibnamefont {Yao}},\ }\bibfield  {title} {\bibinfo {title} {Differentiable {{Quantum Architecture Search}}},\ }\href {https://doi.org/10.1088/2058-9565/ac87cd} {\bibfield  {journal} {\bibinfo  {journal} {Quantum Science and Technology}\ }\textbf {\bibinfo {volume} {7}},\ \bibinfo {pages} {045023} (\bibinfo {year} {2022})},\ \Eprint {https://arxiv.org/abs/2010.08561} {arxiv:2010.08561 [quant-ph]} \BibitemShut {NoStop}%
\bibitem [{\citenamefont {Toulouse}(2006)}]{toulouseAutomaticQuantumComputer2006}%
  \BibitemOpen
  \bibfield  {author} {\bibinfo {author} {\bibfnamefont {M.}~\bibnamefont {Toulouse}},\ }\bibfield  {title} {\bibinfo {title} {Automatic {{Quantum Computer Programming}}: {{A Genetic Programming Approach}}},\ }\href {https://doi.org/10.1007/s10710-006-4866-3} {\bibfield  {journal} {\bibinfo  {journal} {Genetic Programming and Evolvable Machines}\ }\textbf {\bibinfo {volume} {7}},\ \bibinfo {pages} {125} (\bibinfo {year} {2006})}\BibitemShut {NoStop}%
\bibitem [{\citenamefont {Rubinstein}(2001)}]{rubinsteinEvolvingQuantumCircuits2001}%
  \BibitemOpen
  \bibfield  {author} {\bibinfo {author} {\bibfnamefont {B.}~\bibnamefont {Rubinstein}},\ }\bibfield  {title} {\bibinfo {title} {Evolving quantum circuits using genetic programming},\ }in\ \href {https://doi.org/10.1109/CEC.2001.934383} {\emph {\bibinfo {booktitle} {Proceedings of the 2001 {{Congress}} on {{Evolutionary Computation}} ({{IEEE Cat}}. {{No}}.{{01TH8546}})}}},\ Vol.~\bibinfo {volume} {1}\ (\bibinfo  {publisher} {IEEE},\ \bibinfo {address} {Seoul, South Korea},\ \bibinfo {year} {2001})\ pp.\ \bibinfo {pages} {144--151}\BibitemShut {NoStop}%
\bibitem [{\citenamefont {Langdon}\ and\ \citenamefont {Poli}(2002)}]{langdonFoundationsGeneticProgramming2002}%
  \BibitemOpen
  \bibfield  {author} {\bibinfo {author} {\bibfnamefont {W.~B.}\ \bibnamefont {Langdon}}\ and\ \bibinfo {author} {\bibfnamefont {R.}~\bibnamefont {Poli}},\ }\href {https://doi.org/10.1007/978-3-662-04726-2} {\emph {\bibinfo {title} {Foundations of {{Genetic Programming}}}}}\ (\bibinfo  {publisher} {Springer Berlin Heidelberg},\ \bibinfo {address} {Berlin, Heidelberg},\ \bibinfo {year} {2002})\BibitemShut {NoStop}%
\bibitem [{\citenamefont {Kondratyev}(2020)}]{kondratyevNonDifferentiableLearningQuantum2020}%
  \BibitemOpen
  \bibfield  {author} {\bibinfo {author} {\bibfnamefont {A.}~\bibnamefont {Kondratyev}},\ }\bibfield  {title} {\bibinfo {title} {Non-{{Differentiable Learning}} of {{Quantum Circuit Born Machine}} with {{Genetic Algorithm}}},\ }\bibfield  {journal} {\bibinfo  {journal} {SSRN Electronic Journal}\ }\href {https://doi.org/10.2139/ssrn.3569226} {10.2139/ssrn.3569226} (\bibinfo {year} {2020})\BibitemShut {NoStop}%
\bibitem [{\citenamefont {Tang}\ \emph {et~al.}(2021)\citenamefont {Tang}, \citenamefont {Shkolnikov}, \citenamefont {Barron}, \citenamefont {Grimsley}, \citenamefont {Mayhall}, \citenamefont {Barnes},\ and\ \citenamefont {Economou}}]{tangQubitADAPTVQEAdaptiveAlgorithm2021}%
  \BibitemOpen
  \bibfield  {author} {\bibinfo {author} {\bibfnamefont {H.~L.}\ \bibnamefont {Tang}}, \bibinfo {author} {\bibfnamefont {V.}~\bibnamefont {Shkolnikov}}, \bibinfo {author} {\bibfnamefont {G.~S.}\ \bibnamefont {Barron}}, \bibinfo {author} {\bibfnamefont {H.~R.}\ \bibnamefont {Grimsley}}, \bibinfo {author} {\bibfnamefont {N.~J.}\ \bibnamefont {Mayhall}}, \bibinfo {author} {\bibfnamefont {E.}~\bibnamefont {Barnes}},\ and\ \bibinfo {author} {\bibfnamefont {S.~E.}\ \bibnamefont {Economou}},\ }\bibfield  {title} {\bibinfo {title} {Qubit-{{ADAPT-VQE}}: {{An Adaptive Algorithm}} for {{Constructing Hardware-Efficient Ans{\"a}tze}} on a {{Quantum Processor}}},\ }\href {https://doi.org/10.1103/PRXQuantum.2.020310} {\bibfield  {journal} {\bibinfo  {journal} {PRX Quantum}\ }\textbf {\bibinfo {volume} {2}},\ \bibinfo {pages} {020310} (\bibinfo {year} {2021})}\BibitemShut {NoStop}%
\bibitem [{\citenamefont {Ding}\ and\ \citenamefont {Spector}(2023)}]{dingMultiObjectiveEvolutionaryArchitecture2023}%
  \BibitemOpen
  \bibfield  {author} {\bibinfo {author} {\bibfnamefont {L.}~\bibnamefont {Ding}}\ and\ \bibinfo {author} {\bibfnamefont {L.}~\bibnamefont {Spector}},\ }\bibfield  {title} {\bibinfo {title} {Multi-{{Objective Evolutionary Architecture Search}} for {{Parameterized Quantum Circuits}}},\ }\href {https://doi.org/10.3390/e25010093} {\bibfield  {journal} {\bibinfo  {journal} {Entropy}\ }\textbf {\bibinfo {volume} {25}},\ \bibinfo {pages} {93} (\bibinfo {year} {2023})}\BibitemShut {NoStop}%
\bibitem [{\citenamefont {Wolf}\ \emph {et~al.}(2023)\citenamefont {Wolf}, \citenamefont {Ewen},\ and\ \citenamefont {Turkalj}}]{wolfQuantumArchitectureSearch2023}%
  \BibitemOpen
  \bibfield  {author} {\bibinfo {author} {\bibfnamefont {M.-O.}\ \bibnamefont {Wolf}}, \bibinfo {author} {\bibfnamefont {T.}~\bibnamefont {Ewen}},\ and\ \bibinfo {author} {\bibfnamefont {I.}~\bibnamefont {Turkalj}},\ }\bibfield  {title} {\bibinfo {title} {Quantum {{Architecture Search}} for {{Quantum Monte Carlo Integration}} via {{Conditional Parameterized Circuits}} with {{Application}} to {{Finance}}},\ }in\ \href {https://doi.org/10.1109/QCE57702.2023.00070} {\emph {\bibinfo {booktitle} {2023 {{IEEE International Conference}} on {{Quantum Computing}} and {{Engineering}} ({{QCE}})}}},\ Vol.~\bibinfo {volume} {01}\ (\bibinfo {year} {2023})\ pp.\ \bibinfo {pages} {560--570}\BibitemShut {NoStop}%
\bibitem [{\citenamefont {Koza}(1994)}]{kozaGeneticProgrammingMeans1994}%
  \BibitemOpen
  \bibfield  {author} {\bibinfo {author} {\bibfnamefont {J.~R.}\ \bibnamefont {Koza}},\ }\bibfield  {title} {\bibinfo {title} {Genetic programming as a means for programming computers by natural selection},\ }\bibfield  {journal} {\bibinfo  {journal} {Statistics and Computing}\ }\textbf {\bibinfo {volume} {4}},\ \href {https://doi.org/10.1007/BF00175355} {10.1007/BF00175355} (\bibinfo {year} {1994})\BibitemShut {NoStop}%
\bibitem [{\citenamefont {Deb}\ \emph {et~al.}(2002)\citenamefont {Deb}, \citenamefont {Pratap}, \citenamefont {Agarwal},\ and\ \citenamefont {Meyarivan}}]{debFastElitistMultiobjective2002}%
  \BibitemOpen
  \bibfield  {author} {\bibinfo {author} {\bibfnamefont {K.}~\bibnamefont {Deb}}, \bibinfo {author} {\bibfnamefont {A.}~\bibnamefont {Pratap}}, \bibinfo {author} {\bibfnamefont {S.}~\bibnamefont {Agarwal}},\ and\ \bibinfo {author} {\bibfnamefont {T.}~\bibnamefont {Meyarivan}},\ }\bibfield  {title} {\bibinfo {title} {A fast and elitist multiobjective genetic algorithm: {{NSGA-II}}},\ }\href {https://doi.org/10.1109/4235.996017} {\bibfield  {journal} {\bibinfo  {journal} {IEEE Transactions on Evolutionary Computation}\ }\textbf {\bibinfo {volume} {6}},\ \bibinfo {pages} {182} (\bibinfo {year} {2002})}\BibitemShut {NoStop}%
\bibitem [{\citenamefont {Miller}\ and\ \citenamefont {Thomson}(2000)}]{millerCartesianGeneticProgramming2000}%
  \BibitemOpen
  \bibfield  {author} {\bibinfo {author} {\bibfnamefont {J.~F.}\ \bibnamefont {Miller}}\ and\ \bibinfo {author} {\bibfnamefont {P.}~\bibnamefont {Thomson}},\ }\bibfield  {title} {\bibinfo {title} {Cartesian {{Genetic Programming}}},\ }in\ \href {https://doi.org/10.1007/978-3-540-46239-2_9} {\emph {\bibinfo {booktitle} {Genetic {{Programming}}}}},\ \bibinfo {editor} {edited by\ \bibinfo {editor} {\bibfnamefont {R.}~\bibnamefont {Poli}}, \bibinfo {editor} {\bibfnamefont {W.}~\bibnamefont {Banzhaf}}, \bibinfo {editor} {\bibfnamefont {W.~B.}\ \bibnamefont {Langdon}}, \bibinfo {editor} {\bibfnamefont {J.}~\bibnamefont {Miller}}, \bibinfo {editor} {\bibfnamefont {P.}~\bibnamefont {Nordin}},\ and\ \bibinfo {editor} {\bibfnamefont {T.~C.}\ \bibnamefont {Fogarty}}}\ (\bibinfo  {publisher} {Springer},\ \bibinfo {address} {Berlin, Heidelberg},\ \bibinfo {year} {2000})\ pp.\ \bibinfo {pages} {121--132}\BibitemShut {NoStop}%
\bibitem [{\citenamefont {Barnes}(2020)}]{barnesGeneticEvolutionQuantum2020}%
  \BibitemOpen
  \bibfield  {author} {\bibinfo {author} {\bibfnamefont {K.}~\bibnamefont {Barnes}},\ }\bibfield  {title} {\bibinfo {title} {The {{Genetic Evolution}} of {{Quantum Programs Using The ZX-Calculus}}}} (\bibinfo {year} {2020})\BibitemShut {NoStop}%
\bibitem [{\citenamefont {Coecke}\ and\ \citenamefont {Duncan}(2008)}]{coeckeInteractingQuantumObservables2008}%
  \BibitemOpen
  \bibfield  {author} {\bibinfo {author} {\bibfnamefont {B.}~\bibnamefont {Coecke}}\ and\ \bibinfo {author} {\bibfnamefont {R.}~\bibnamefont {Duncan}},\ }\bibfield  {title} {\bibinfo {title} {Interacting {{Quantum Observables}}},\ }in\ \href {https://doi.org/10.1007/978-3-540-70583-3_25} {\emph {\bibinfo {booktitle} {Automata, {{Languages}} and {{Programming}}}}},\ Vol.\ \bibinfo {volume} {5126},\ \bibinfo {editor} {edited by\ \bibinfo {editor} {\bibfnamefont {L.}~\bibnamefont {Aceto}}, \bibinfo {editor} {\bibfnamefont {I.}~\bibnamefont {Damg{\aa}rd}}, \bibinfo {editor} {\bibfnamefont {L.~A.}\ \bibnamefont {Goldberg}}, \bibinfo {editor} {\bibfnamefont {M.~M.}\ \bibnamefont {Halld{\'o}rsson}}, \bibinfo {editor} {\bibfnamefont {A.}~\bibnamefont {Ing{\'o}lfsd{\'o}ttir}},\ and\ \bibinfo {editor} {\bibfnamefont {I.}~\bibnamefont {Walukiewicz}}}\ (\bibinfo  {publisher} {Springer Berlin Heidelberg},\ \bibinfo {address} {Berlin, Heidelberg},\ \bibinfo {year} {2008})\ pp.\ \bibinfo {pages} {298--310}\BibitemShut {NoStop}%
\bibitem [{\citenamefont {Chancellor}\ \emph {et~al.}(2023)\citenamefont {Chancellor}, \citenamefont {Kissinger}, \citenamefont {Roffe}, \citenamefont {Zohren},\ and\ \citenamefont {Horsman}}]{chancellorGraphicalStructuresDesign2023}%
  \BibitemOpen
  \bibfield  {author} {\bibinfo {author} {\bibfnamefont {N.}~\bibnamefont {Chancellor}}, \bibinfo {author} {\bibfnamefont {A.}~\bibnamefont {Kissinger}}, \bibinfo {author} {\bibfnamefont {J.}~\bibnamefont {Roffe}}, \bibinfo {author} {\bibfnamefont {S.}~\bibnamefont {Zohren}},\ and\ \bibinfo {author} {\bibfnamefont {D.}~\bibnamefont {Horsman}},\ }\bibfield  {title} {\bibinfo {title} {Graphical {{Structures}} for {{Design}} and {{Verification}} of {{Quantum Error Correction}}},\ }\href {https://doi.org/10.1088/2058-9565/acf157} {\bibfield  {journal} {\bibinfo  {journal} {Quantum Science and Technology}\ }\textbf {\bibinfo {volume} {8}},\ \bibinfo {pages} {045028} (\bibinfo {year} {2023})},\ \Eprint {https://arxiv.org/abs/1611.08012} {arxiv:1611.08012 [quant-ph]} \BibitemShut {NoStop}%
\bibitem [{\citenamefont {Duncan}\ and\ \citenamefont {Perdrix}(2010)}]{duncanRewritingMeasurementBasedQuantum2010}%
  \BibitemOpen
  \bibfield  {author} {\bibinfo {author} {\bibfnamefont {R.}~\bibnamefont {Duncan}}\ and\ \bibinfo {author} {\bibfnamefont {S.}~\bibnamefont {Perdrix}},\ }\bibfield  {title} {\bibinfo {title} {Rewriting {{Measurement-Based Quantum Computations}} with {{Generalised Flow}}},\ }in\ \href {https://doi.org/10.1007/978-3-642-14162-1_24} {\emph {\bibinfo {booktitle} {Automata, {{Languages}} and {{Programming}}}}},\ \bibinfo {editor} {edited by\ \bibinfo {editor} {\bibfnamefont {S.}~\bibnamefont {Abramsky}}, \bibinfo {editor} {\bibfnamefont {C.}~\bibnamefont {Gavoille}}, \bibinfo {editor} {\bibfnamefont {C.}~\bibnamefont {Kirchner}}, \bibinfo {editor} {\bibfnamefont {F.}~\bibnamefont {{Meyer auf der Heide}}},\ and\ \bibinfo {editor} {\bibfnamefont {P.~G.}\ \bibnamefont {Spirakis}}}\ (\bibinfo  {publisher} {Springer},\ \bibinfo {address} {Berlin, Heidelberg},\ \bibinfo {year} {2010})\ pp.\ \bibinfo {pages} {285--296}\BibitemShut {NoStop}%
\bibitem [{\citenamefont {Gogioso}\ and\ \citenamefont {Yeung}(2023)}]{gogiosoAnnealingOptimisationMixed2023}%
  \BibitemOpen
  \bibfield  {author} {\bibinfo {author} {\bibfnamefont {S.}~\bibnamefont {Gogioso}}\ and\ \bibinfo {author} {\bibfnamefont {R.}~\bibnamefont {Yeung}},\ }\bibfield  {title} {\bibinfo {title} {Annealing {{Optimisation}} of {{Mixed ZX Phase Circuits}}},\ }\href {https://doi.org/10.4204/EPTCS.394.20} {\bibfield  {journal} {\bibinfo  {journal} {Electronic Proceedings in Theoretical Computer Science}\ }\textbf {\bibinfo {volume} {394}},\ \bibinfo {pages} {415} (\bibinfo {year} {2023})}\BibitemShut {NoStop}%
\bibitem [{\citenamefont {Duncan}\ \emph {et~al.}(2020)\citenamefont {Duncan}, \citenamefont {Kissinger}, \citenamefont {Perdrix},\ and\ \citenamefont {{van de Wetering}}}]{duncanGraphtheoreticSimplificationQuantum2020}%
  \BibitemOpen
  \bibfield  {author} {\bibinfo {author} {\bibfnamefont {R.}~\bibnamefont {Duncan}}, \bibinfo {author} {\bibfnamefont {A.}~\bibnamefont {Kissinger}}, \bibinfo {author} {\bibfnamefont {S.}~\bibnamefont {Perdrix}},\ and\ \bibinfo {author} {\bibfnamefont {J.}~\bibnamefont {{van de Wetering}}},\ }\bibfield  {title} {\bibinfo {title} {Graph-theoretic {{Simplification}} of {{Quantum Circuits}} with the {{ZX-calculus}}},\ }\href {https://doi.org/10.22331/q-2020-06-04-279} {\bibfield  {journal} {\bibinfo  {journal} {Quantum}\ }\textbf {\bibinfo {volume} {4}},\ \bibinfo {pages} {279} (\bibinfo {year} {2020})},\ \Eprint {https://arxiv.org/abs/1902.03178} {arxiv:1902.03178 [quant-ph]} \BibitemShut {NoStop}%
\bibitem [{\citenamefont {Kissinger}\ and\ \citenamefont {{van de Wetering}}(2020{\natexlab{a}})}]{kissingerReducingNumberNonClifford2020}%
  \BibitemOpen
  \bibfield  {author} {\bibinfo {author} {\bibfnamefont {A.}~\bibnamefont {Kissinger}}\ and\ \bibinfo {author} {\bibfnamefont {J.}~\bibnamefont {{van de Wetering}}},\ }\bibfield  {title} {\bibinfo {title} {Reducing the number of non-{{Clifford}} gates in quantum circuits},\ }\href {https://doi.org/10.1103/PhysRevA.102.022406} {\bibfield  {journal} {\bibinfo  {journal} {Physical Review A}\ }\textbf {\bibinfo {volume} {102}},\ \bibinfo {pages} {022406} (\bibinfo {year} {2020}{\natexlab{a}})}\BibitemShut {NoStop}%
\bibitem [{\citenamefont {Backens}\ \emph {et~al.}(2021)\citenamefont {Backens}, \citenamefont {{Miller-Bakewell}}, \citenamefont {de~Felice}, \citenamefont {Lobski},\ and\ \citenamefont {van~de Wetering}}]{backensThereBackAgain2021}%
  \BibitemOpen
  \bibfield  {author} {\bibinfo {author} {\bibfnamefont {M.}~\bibnamefont {Backens}}, \bibinfo {author} {\bibfnamefont {H.}~\bibnamefont {{Miller-Bakewell}}}, \bibinfo {author} {\bibfnamefont {G.}~\bibnamefont {de~Felice}}, \bibinfo {author} {\bibfnamefont {L.}~\bibnamefont {Lobski}},\ and\ \bibinfo {author} {\bibfnamefont {J.}~\bibnamefont {van~de Wetering}},\ }\bibfield  {title} {\bibinfo {title} {There and back again: {{A}} circuit extraction tale},\ }\href {https://doi.org/10.22331/q-2021-03-25-421} {\bibfield  {journal} {\bibinfo  {journal} {Quantum}\ }\textbf {\bibinfo {volume} {5}},\ \bibinfo {pages} {421} (\bibinfo {year} {2021})}\BibitemShut {NoStop}%
\bibitem [{\citenamefont {{van de Wetering}}(2020)}]{vandeweteringZXcalculusWorkingQuantum2020}%
  \BibitemOpen
  \bibfield  {author} {\bibinfo {author} {\bibfnamefont {J.}~\bibnamefont {{van de Wetering}}},\ }\href {https://doi.org/10.48550/arXiv.2012.13966} {\bibinfo {title} {{{ZX-calculus}} for the working quantum computer scientist}} (\bibinfo {year} {2020}),\ \Eprint {https://arxiv.org/abs/2012.13966} {arxiv:2012.13966 [quant-ph]} \BibitemShut {NoStop}%
\bibitem [{\citenamefont {Coecke}\ and\ \citenamefont {Kissinger}(2017)}]{coeckePicturingQuantumProcesses2017}%
  \BibitemOpen
  \bibfield  {author} {\bibinfo {author} {\bibfnamefont {B.}~\bibnamefont {Coecke}}\ and\ \bibinfo {author} {\bibfnamefont {A.}~\bibnamefont {Kissinger}},\ }\href {https://doi.org/10.1017/9781316219317} {\emph {\bibinfo {title} {Picturing {{Quantum Processes}}: {{A First Course}} in {{Quantum Theory}} and {{Diagrammatic Reasoning}}}}}\ (\bibinfo  {publisher} {Cambridge University Press},\ \bibinfo {address} {Cambridge},\ \bibinfo {year} {2017})\BibitemShut {NoStop}%
\bibitem [{\citenamefont {Coecke}\ and\ \citenamefont {Gogioso}(2022)}]{coeckeQuantumPictures2022}%
  \BibitemOpen
  \bibfield  {author} {\bibinfo {author} {\bibfnamefont {B.}~\bibnamefont {Coecke}}\ and\ \bibinfo {author} {\bibfnamefont {S.}~\bibnamefont {Gogioso}},\ }\href@noop {} {\emph {\bibinfo {title} {Quantum in {{Pictures}}}}}\ (\bibinfo  {publisher} {Quantinuum},\ \bibinfo {year} {2022})\BibitemShut {NoStop}%
\bibitem [{\citenamefont {Hazewinkel}(1996)}]{hazewinkelHandbookAlgebraVolume1996}%
  \BibitemOpen
  \bibfield  {author} {\bibinfo {author} {\bibfnamefont {M.}~\bibnamefont {Hazewinkel}},\ }\href@noop {} {\emph {\bibinfo {title} {Handbook of Algebra. {{Volume}} 1}}}\ (\bibinfo  {publisher} {Elsevier},\ \bibinfo {address} {Amsterdam},\ \bibinfo {year} {1996})\BibitemShut {NoStop}%
\bibitem [{\citenamefont {Piedeleu}\ and\ \citenamefont {Zanasi}(2023)}]{piedeleuIntroductionStringDiagrams2023}%
  \BibitemOpen
  \bibfield  {author} {\bibinfo {author} {\bibfnamefont {R.}~\bibnamefont {Piedeleu}}\ and\ \bibinfo {author} {\bibfnamefont {F.}~\bibnamefont {Zanasi}},\ }\href@noop {} {\bibinfo {title} {An {{Introduction}} to {{String Diagrams}} for {{Computer Scientists}}}} (\bibinfo {year} {2023}),\ \Eprint {https://arxiv.org/abs/2305.08768} {arxiv:2305.08768 [cs]} \BibitemShut {NoStop}%
\bibitem [{\citenamefont {Hinze}\ and\ \citenamefont {Marsden}(2023)}]{hinzeIntroducingStringDiagrams2023}%
  \BibitemOpen
  \bibfield  {author} {\bibinfo {author} {\bibfnamefont {R.}~\bibnamefont {Hinze}}\ and\ \bibinfo {author} {\bibfnamefont {D.}~\bibnamefont {Marsden}},\ }\href {https://doi.org/10.1017/9781009317825} {\emph {\bibinfo {title} {Introducing {{String Diagrams}}: {{The Art}} of {{Category Theory}}}}}\ (\bibinfo  {publisher} {Cambridge University Press},\ \bibinfo {address} {Cambridge},\ \bibinfo {year} {2023})\BibitemShut {NoStop}%
\bibitem [{\citenamefont {Kissinger}\ and\ \citenamefont {{van de Wetering}}(2020{\natexlab{b}})}]{kissingerPyZXLargeScale2020}%
  \BibitemOpen
  \bibfield  {author} {\bibinfo {author} {\bibfnamefont {A.}~\bibnamefont {Kissinger}}\ and\ \bibinfo {author} {\bibfnamefont {J.}~\bibnamefont {{van de Wetering}}},\ }\bibfield  {title} {\bibinfo {title} {{{PyZX}}: {{Large Scale Automated Diagrammatic Reasoning}}},\ }\href {https://doi.org/10.4204/EPTCS.318.14} {\bibfield  {journal} {\bibinfo  {journal} {Electronic Proceedings in Theoretical Computer Science}\ }\textbf {\bibinfo {volume} {318}},\ \bibinfo {pages} {229} (\bibinfo {year} {2020}{\natexlab{b}})},\ \Eprint {https://arxiv.org/abs/1904.04735} {arxiv:1904.04735 [quant-ph]} \BibitemShut {NoStop}%
\bibitem [{\citenamefont {Meurer}\ \emph {et~al.}(2017)\citenamefont {Meurer}, \citenamefont {Smith}, \citenamefont {Paprocki}, \citenamefont {{\v C}ert{\'i}k}, \citenamefont {Kirpichev}, \citenamefont {Rocklin}, \citenamefont {Kumar}, \citenamefont {Ivanov}, \citenamefont {Moore}, \citenamefont {Singh}, \citenamefont {Rathnayake}, \citenamefont {Vig}, \citenamefont {Granger}, \citenamefont {Muller}, \citenamefont {Bonazzi}, \citenamefont {Gupta}, \citenamefont {Vats}, \citenamefont {Johansson}, \citenamefont {Pedregosa}, \citenamefont {Curry}, \citenamefont {Terrel}, \citenamefont {Rou{\v c}ka}, \citenamefont {Saboo}, \citenamefont {Fernando}, \citenamefont {Kulal}, \citenamefont {Cimrman},\ and\ \citenamefont {Scopatz}}]{10.7717/peerj-cs.103}%
  \BibitemOpen
  \bibfield  {author} {\bibinfo {author} {\bibfnamefont {A.}~\bibnamefont {Meurer}}, \bibinfo {author} {\bibfnamefont {C.~P.}\ \bibnamefont {Smith}}, \bibinfo {author} {\bibfnamefont {M.}~\bibnamefont {Paprocki}}, \bibinfo {author} {\bibfnamefont {O.}~\bibnamefont {{\v C}ert{\'i}k}}, \bibinfo {author} {\bibfnamefont {S.~B.}\ \bibnamefont {Kirpichev}}, \bibinfo {author} {\bibfnamefont {M.}~\bibnamefont {Rocklin}}, \bibinfo {author} {\bibfnamefont {{\relax Am}.}~\bibnamefont {Kumar}}, \bibinfo {author} {\bibfnamefont {S.}~\bibnamefont {Ivanov}}, \bibinfo {author} {\bibfnamefont {J.~K.}\ \bibnamefont {Moore}}, \bibinfo {author} {\bibfnamefont {S.}~\bibnamefont {Singh}} \emph {et~al.},\ }\bibfield  {title} {\bibinfo {title} {{{SymPy}}: Symbolic computing in {{Python}}},\ }\href {https://doi.org/10.7717/peerj-cs.103} {\bibfield  {journal} {\bibinfo  {journal} {PeerJ Computer Science}\ }\textbf {\bibinfo {volume} {3}},\ \bibinfo {pages} {e103} (\bibinfo {year} {2017})}\BibitemShut {NoStop}%
\bibitem [{\citenamefont {Bergholm}\ \emph {et~al.}(2022)\citenamefont {Bergholm}, \citenamefont {Izaac}, \citenamefont {Schuld}, \citenamefont {Gogolin}, \citenamefont {Ahmed}, \citenamefont {Ajith}, \citenamefont {Alam}, \citenamefont {{Alonso-Linaje}}, \citenamefont {AkashNarayanan}, \citenamefont {Asadi}, \citenamefont {Arrazola}, \citenamefont {Azad}, \citenamefont {Banning}, \citenamefont {Blank}, \citenamefont {Bromley}, \citenamefont {Cordier}, \citenamefont {Ceroni}, \citenamefont {Delgado}, \citenamefont {Di~Matteo}, \citenamefont {Dusko}, \citenamefont {Garg}, \citenamefont {Guala}, \citenamefont {Hayes}, \citenamefont {Hill}, \citenamefont {Ijaz}, \citenamefont {Isacsson}, \citenamefont {Ittah}, \citenamefont {Jahangiri}, \citenamefont {Jain}, \citenamefont {Jiang}, \citenamefont {Khandelwal}, \citenamefont {Kottmann}, \citenamefont {Lang}, \citenamefont {Lee}, \citenamefont {Loke}, \citenamefont {Lowe}, \citenamefont {McKiernan}, \citenamefont {Meyer}, \citenamefont {{Monta{\~n}ez-Barrera}}, \citenamefont {Moyard}, \citenamefont {Niu}, \citenamefont {O'Riordan}, \citenamefont {Oud}, \citenamefont {Panigrahi}, \citenamefont {Park}, \citenamefont {Polatajko}, \citenamefont {Quesada}, \citenamefont {Roberts}, \citenamefont {S{\'a}}, \citenamefont {Schoch}, \citenamefont {Shi}, \citenamefont {Shu}, \citenamefont {Sim}, \citenamefont {Singh}, \citenamefont {Strandberg}, \citenamefont {Soni}, \citenamefont {Sz{\'a}va}, \citenamefont {Thabet}, \citenamefont {{Vargas-Hern{\'a}ndez}}, \citenamefont {Vincent}, \citenamefont {Vitucci}, \citenamefont {Weber}, \citenamefont {Wierichs}, \citenamefont {Wiersema}, \citenamefont {Willmann}, \citenamefont {Wong}, \citenamefont {Zhang},\ and\ \citenamefont {Killoran}}]{bergholmPennyLaneAutomaticDifferentiation2022a}%
  \BibitemOpen
  \bibfield  {author} {\bibinfo {author} {\bibfnamefont {V.}~\bibnamefont {Bergholm}}, \bibinfo {author} {\bibfnamefont {J.}~\bibnamefont {Izaac}}, \bibinfo {author} {\bibfnamefont {M.}~\bibnamefont {Schuld}}, \bibinfo {author} {\bibfnamefont {C.}~\bibnamefont {Gogolin}}, \bibinfo {author} {\bibfnamefont {S.}~\bibnamefont {Ahmed}}, \bibinfo {author} {\bibfnamefont {V.}~\bibnamefont {Ajith}}, \bibinfo {author} {\bibfnamefont {M.~S.}\ \bibnamefont {Alam}}, \bibinfo {author} {\bibfnamefont {G.}~\bibnamefont {{Alonso-Linaje}}}, \bibinfo {author} {\bibfnamefont {B.}~\bibnamefont {AkashNarayanan}}, \bibinfo {author} {\bibfnamefont {A.}~\bibnamefont {Asadi}} \emph {et~al.},\ }\href {https://doi.org/10.48550/arXiv.1811.04968} {\bibinfo {title} {{{PennyLane}}: {{Automatic}} differentiation of hybrid quantum-classical computations}} (\bibinfo {year} {2022}),\ \Eprint {https://arxiv.org/abs/1811.04968} {arxiv:1811.04968 [physics, physics:quant-ph]} \BibitemShut {NoStop}%
\bibitem [{\citenamefont {{DeepMind}}\ \emph {et~al.}(2020)\citenamefont {{DeepMind}}, \citenamefont {Babuschkin}, \citenamefont {Baumli}, \citenamefont {Bell}, \citenamefont {Bhupatiraju}, \citenamefont {Bruce}, \citenamefont {Buchlovsky}, \citenamefont {Budden}, \citenamefont {Cai}, \citenamefont {Clark}, \citenamefont {Danihelka}, \citenamefont {Dedieu}, \citenamefont {Fantacci}, \citenamefont {Godwin}, \citenamefont {Jones}, \citenamefont {Hemsley}, \citenamefont {Hennigan}, \citenamefont {Hessel}, \citenamefont {Hou}, \citenamefont {Kapturowski}, \citenamefont {Keck}, \citenamefont {Kemaev}, \citenamefont {King}, \citenamefont {Kunesch}, \citenamefont {Martens}, \citenamefont {Merzic}, \citenamefont {Mikulik}, \citenamefont {Norman}, \citenamefont {Papamakarios}, \citenamefont {Quan}, \citenamefont {Ring}, \citenamefont {Ruiz}, \citenamefont {Sanchez}, \citenamefont {Sartran}, \citenamefont {Schneider}, \citenamefont {Sezener}, \citenamefont {Spencer}, \citenamefont {Srinivasan}, \citenamefont {Stanojevi{\'c}}, \citenamefont {Stokowiec}, \citenamefont {Wang}, \citenamefont {Zhou},\ and\ \citenamefont {Viola}}]{deepmind2020jax}%
  \BibitemOpen
  \bibfield  {author} {\bibinfo {author} {\bibnamefont {{DeepMind}}}, \bibinfo {author} {\bibfnamefont {I.}~\bibnamefont {Babuschkin}}, \bibinfo {author} {\bibfnamefont {K.}~\bibnamefont {Baumli}}, \bibinfo {author} {\bibfnamefont {A.}~\bibnamefont {Bell}}, \bibinfo {author} {\bibfnamefont {S.}~\bibnamefont {Bhupatiraju}}, \bibinfo {author} {\bibfnamefont {J.}~\bibnamefont {Bruce}}, \bibinfo {author} {\bibfnamefont {P.}~\bibnamefont {Buchlovsky}}, \bibinfo {author} {\bibfnamefont {D.}~\bibnamefont {Budden}}, \bibinfo {author} {\bibfnamefont {T.}~\bibnamefont {Cai}}, \bibinfo {author} {\bibfnamefont {A.}~\bibnamefont {Clark}} \emph {et~al.},\ }\href@noop {} {\bibinfo {title} {The {{DeepMind JAX Ecosystem}}}} (\bibinfo {year} {2020})\BibitemShut {NoStop}%
\bibitem [{\citenamefont {Spall}(1998)}]{spallOverviewSimultaneousPerturbation1998}%
  \BibitemOpen
  \bibfield  {author} {\bibinfo {author} {\bibfnamefont {J.~C.}\ \bibnamefont {Spall}},\ }\bibfield  {title} {\bibinfo {title} {An {{Overview}} of the {{Simultaneous Perturbation Method}} for {{Efficient Optimization}}},\ }\href@noop {} {\bibfield  {journal} {\bibinfo  {journal} {Johns Hopkins apl technical digest}\ }\textbf {\bibinfo {volume} {19}},\ \bibinfo {pages} {11} (\bibinfo {year} {1998})}\BibitemShut {NoStop}%
\bibitem [{\citenamefont {Schuld}\ \emph {et~al.}(2019)\citenamefont {Schuld}, \citenamefont {Bergholm}, \citenamefont {Gogolin}, \citenamefont {Izaac},\ and\ \citenamefont {Killoran}}]{schuldEvaluatingAnalyticGradients2019}%
  \BibitemOpen
  \bibfield  {author} {\bibinfo {author} {\bibfnamefont {M.}~\bibnamefont {Schuld}}, \bibinfo {author} {\bibfnamefont {V.}~\bibnamefont {Bergholm}}, \bibinfo {author} {\bibfnamefont {C.}~\bibnamefont {Gogolin}}, \bibinfo {author} {\bibfnamefont {J.}~\bibnamefont {Izaac}},\ and\ \bibinfo {author} {\bibfnamefont {N.}~\bibnamefont {Killoran}},\ }\bibfield  {title} {\bibinfo {title} {Evaluating analytic gradients on quantum hardware},\ }\href {https://doi.org/10.1103/PhysRevA.99.032331} {\bibfield  {journal} {\bibinfo  {journal} {Physical Review A}\ }\textbf {\bibinfo {volume} {99}},\ \bibinfo {pages} {032331} (\bibinfo {year} {2019})},\ \Eprint {https://arxiv.org/abs/1811.11184} {arxiv:1811.11184 [quant-ph]} \BibitemShut {NoStop}%
\bibitem [{\citenamefont {Schuld}\ and\ \citenamefont {Killoran}(2022)}]{schuldQuantumAdvantageRight2022}%
  \BibitemOpen
  \bibfield  {author} {\bibinfo {author} {\bibfnamefont {M.}~\bibnamefont {Schuld}}\ and\ \bibinfo {author} {\bibfnamefont {N.}~\bibnamefont {Killoran}},\ }\bibfield  {title} {\bibinfo {title} {Is {{Quantum Advantage}} the {{Right Goal}} for {{Quantum Machine Learning}}?},\ }\href {https://doi.org/10.1103/PRXQuantum.3.030101} {\bibfield  {journal} {\bibinfo  {journal} {PRX Quantum}\ }\textbf {\bibinfo {volume} {3}},\ \bibinfo {pages} {030101} (\bibinfo {year} {2022})}\BibitemShut {NoStop}%
\end{thebibliography}%

\end{document}